\documentclass[aps,prl,article,twocolumn,citeautoscript,footinbib,superscriptaddress,nobalancelastpage]{revtex4-1} \synctex=1
\usepackage{amsmath,amssymb,bm,bbm}
\usepackage{amsthm}
\usepackage{physics}
\usepackage{float}
\usepackage{graphicx}
\usepackage{amsmath,amsfonts,tikz}
\usepackage{graphicx,xcolor,multirow}
\usepackage{hyperref}
\usepackage[section]{placeins}
\usepackage[mathscr]{euscript}
\usepackage[toc,page]{appendix}
\hypersetup{
    pdfstartview={FitH},    
    colorlinks=true,       
    linkcolor=blue,          
    citecolor=blue,        
    filecolor=magenta,      
    urlcolor=blue           
}

\usepackage{pdfpages}
\usepackage{pgffor}
\makeatletter
\AtBeginDocument{\let\LS@rot\@undefined}
\makeatother

\usepackage{pifont}

\usepackage{amsfonts}
\usepackage{upgreek}
\usepackage{slashed}
\usepackage{latexsym}
\usepackage{mathrsfs}
\usepackage{tikz-feynman}
\usepackage[usestackEOL]{stackengine}
\usepackage{natbib}
\begin{document}
\title{Emergent glassy behavior in a kagome Rydberg atom array}

\author{Zheng Yan}
\affiliation{Department of Physics and HKU-UCAS Joint Institute of Theoretical and Computational Physics, The University of Hong Kong, Pokfulam Road, Hong Kong SAR, China}

\author{Yan-Cheng Wang}
\affiliation{Zhongfa Aviation Institute of Beihang University, Hangzhou 310023, China}

\author{Rhine Samajdar}
\affiliation{Department of Physics, Princeton University, Princeton, NJ 08544, USA}
\affiliation{Princeton Center for Theoretical Science, Princeton University, Princeton, NJ 08544, USA}

\author{Subir Sachdev}
\email{sachdev@g.harvard.edu}
\affiliation{Department of Physics, Harvard University, Cambridge, MA 02138, USA}

\author{Zi Yang Meng}
\email{zymeng@hku.hk}
\affiliation{Department of Physics and HKU-UCAS Joint Institute of Theoretical and Computational Physics, The University of Hong Kong, Pokfulam Road, Hong Kong SAR, China}
\date{\today}

\begin{abstract}
We present large-scale quantum Monte Carlo simulation results on a realistic Hamiltonian of kagome-lattice Rydberg atom arrays. Although the system has no intrinsic disorder, intriguingly, our analyses of static and dynamic properties on large system sizes reveal \textit{emergent} glassy behavior in a region of parameter space located between two valence bond solid phases. The extent of this glassy region is demarcated using the Edwards-Anderson order parameter, and its phase transitions to the two proximate valence bond solids---as well as the crossover towards a trivial paramagnetic phase---are identified. We demonstrate the intrinsically slow (imaginary) time dynamics deep inside the glassy phase and discuss experimental considerations for detecting such a quantum disordered phase with numerous nearly degenerate local minima. Our proposal paves a new route to the study of real-time glassy phenomena and highlights the potential for quantum simulation of a distinct phase of quantum matter beyond solids and liquids in current-generation Rydberg platforms.
\end{abstract}

\maketitle

\textit{Introduction.---}Over the last decade, quantum simulators based on programmable Rydberg atom arrays \cite{endres2016atom,bernien2017probing,Browaeys.2020,morgado2021quantum,bluvstein2022quantum} have emerged as powerful platforms for the investigation of highly correlated quantum matter. These systems have opened up new avenues to study interesting many-body states \cite{de2019observation,Ebadi.2021,scholl2021quantum,Semeghini21}, quantum dynamics \cite{turner2018weak,keesling2019quantum,bluvstein2021controlling}, gauge theories \cite{Celi1,Surace1,qudit,IGT}, and even combinatorial optimization problems \cite{pichler2018quantum,Ebadi2022,yan2021sweeping}.

An especially promising direction that has recently attracted much attention is the simulation of quantum phases of matter in these tunable atomic setups. Such phases and the transitions between them have been intensely studied for Rydberg atoms arrayed in one spatial dimension \cite{samajdar2018numerical,whitsitt2018quantum,PhysRevLett.122.017205,chepiga2021kibble} as well as in various two-dimensional geometries, including on the square \cite{Samajdar_2020,PhysRevLett.126.170603,kalinowski2021bulk,Kim.2021,orourke2022entanglement}, triangular \cite{li2022quantum}, honeycomb \cite{honeycomb,trimer}, kagome \cite{Samajdar.2021}, and ruby \cite{Verresen.2020, hannes_dynamical,zhai} lattices. In particular, Ref.~\onlinecite{Samajdar.2021} identified an intriguing  highly correlated regime in the phase diagram of the kagome-lattice Rydberg atom array characterized by a lack of symmetry-breaking solid order and a large entanglement entropy. The correlations in this region were found to be ``liquid-like'' in that the density of excitations is limited by the strong Rydberg-Rydberg interactions (as opposed to a weakly interacting gas wherein the laser drive induces independent atomic excitations). Mapping this system to a quantum dimer model \cite{moessner2011quantum} on the triangular lattice \cite{moessner2001ising,roychowdhury2015z,ZY2022loop} raises the possibility of a spin liquid phase with $\mathbb{Z}_2$ topological order \cite{Samajdar.2021,yan2022triangular}. However, since the precise microscopic interactions differ between the Rydberg \cite{Samajdar.2021} and dimer \cite{yan2022triangular} models, it is crucial to \textit{independently} establish the properties of the former in the thermodynamic limit. This poses a challenging problem for numerical techniques such as the density-matrix renormalization group (DMRG), which is not only hindered by the geometrically frustrated and long-ranged nature of the Hamiltonian in Eq.~\eqref{eq:eq1} but is also limited to relatively small system sizes on cylinders.

\begin{figure}[tb]
\centering
\includegraphics[width=\linewidth]{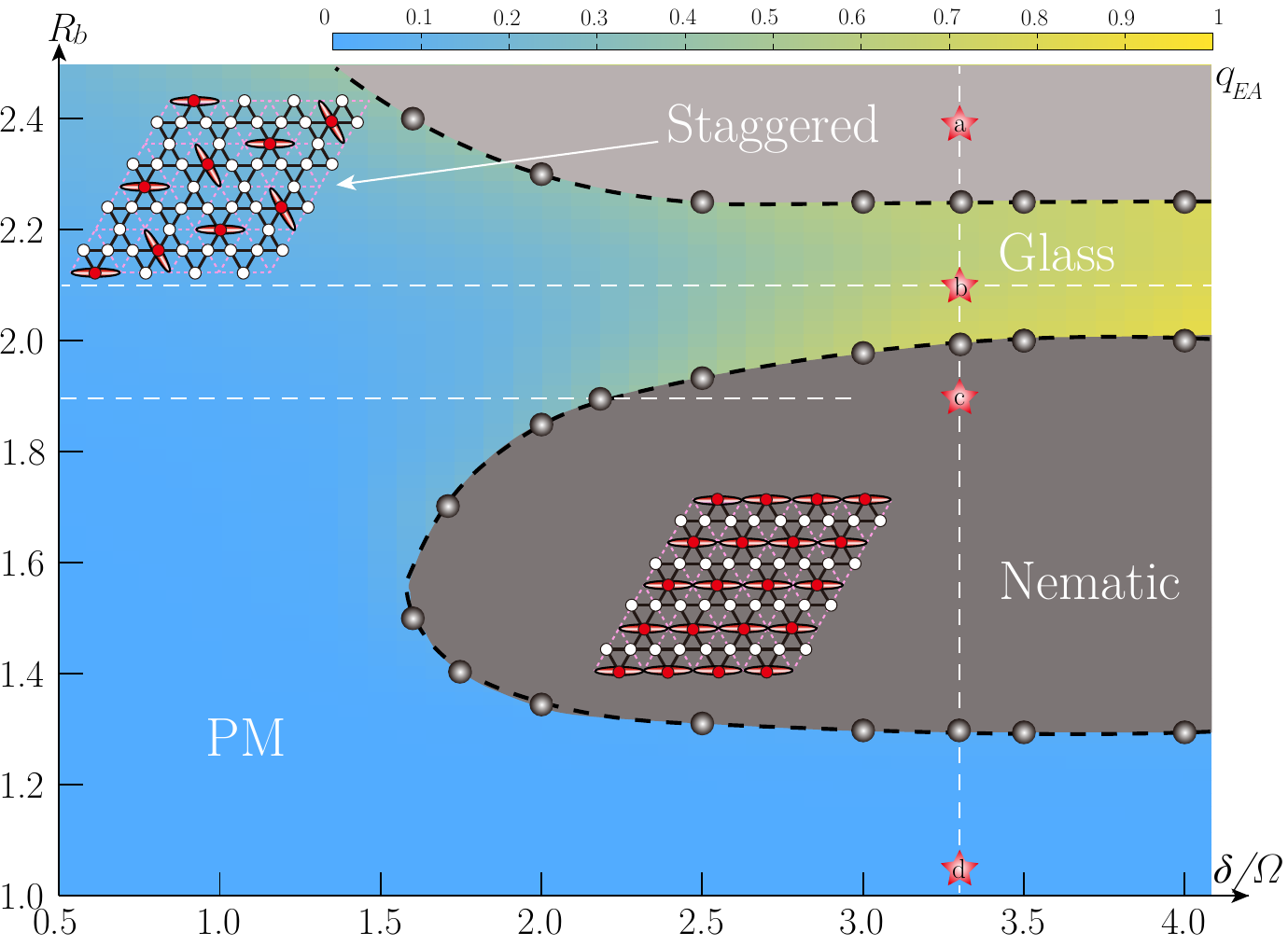}
\caption{Phase diagram, spanned by the $\delta/\Omega$ and $R_b$ axes, obtained from QMC simulations. The four stars mark the points in each phase for which the equal-time dimer structure factor is presented in Fig.~\ref{Fig2}. The white dashed lines indicate the cuts in parameter space along which the quantum phase transitions are studied in Figs.~\ref{Fig3} and S2. The colored background shows the Edwards-Anderson order parameter as obtained in Fig.~\ref{Fig3}(c), with the colorbar on top denoting the scale of $q^{}_{\textsc{ea}}$. 
The two insets schematically sketch the two crystalline phases (nematic and staggered). Each  red (white) circle representing an atom in the Rydberg (ground) state on the kagome lattice [black] can be mapped to the presence (absence) of a dimer on the medial triangular lattice [white].
}
    \label{Fig1}
\end{figure}

In light of the situation, here, we overcome this obstacle and present large-scale quantum Monte Carlo (QMC) simulation results on the realistic Hamiltonian of  kagome-lattice Rydberg arrays in Eq.~\eqref{eq:eq1} below. Surprisingly, even though the Hamiltonian is translationally invariant and has no disorder, our unbiased numerical results for large system sizes and dynamic and static data reveal \textit{emergent} glassy behavior~\cite{knap2019} in the region  located between the so-called ``nematic'' \cite{roychowdhury2015z,Plat2015z2} and ``staggered'' \cite{RMSLS01} valence bond solid (VBS) phases \cite{Samajdar.2021}. 
Although disorder-free glassiness was observed in previous results obtained in extended Heisenberg models \cite{angelone2016superglass}, it is novel in the context of the realistic Rydberg arrays.
Moreover, we emphasize that unlike previous work that identified glassy behavior in open dissipative Rydberg gases~\cite{lesanovsky2013kinetic,carlos2018}, our findings here apply to an isolated closed quantum system. We utilize the Edwards-Anderson order parameter to map out the extent of the glassy region in the phase diagram. Furthermore, the phase transitions between the glassy phase and the two valence bond phases as well as the crossover towards the paramagnetic phase are identified. Our results highlight the intrinsically slow (imaginary) time dynamics deep inside the glassy phase, and we suggest experimental protocols to detect such a quantum disordered phase with numerous nearly degenerate local minima in its energy landscape.

\textit{Rydberg Hamiltonian on the kagome lattice.---}We investigate the following realistic Hamiltonian describing Rydberg arrays on the kagome lattice,
\begin{alignat}{1}
  H=&\sum_{i=1}^N \left[\frac\Omega 2\left(
  \left|g\right>_i\left<r\right| + \left|r\right>_i\left<g\right|\right)-\delta\left|r\right>_i\left<r\right|\right]\nonumber \\
  +&\sum_{i,j=1}^N \frac {V_{ij}} {2} \left(\left|r\right>_i\left<r\right|\otimes\left|r\right>_j\left<r\right|\right),
\label{eq:eq1}
\end{alignat}
where the sum on $i$ runs over all $N$ sites of the kagome lattice. The ket $\left|g\right>$ ($\left|r\right>$) represents the ground (Rydberg) state, while $\Omega$ ($\delta$) stands for the Rabi frequency (detuning) of the laser drive, which can be mapped to a transverse (longitudinal) field in the language of quantum Ising model. The repulsive interaction is of the van der Waals form $V_{ij}$\,$=$\,$\Omega R_b^6/R_{ij}^6$, where $R_{ij}$ is the distance between the sites $i$ and $j$, and $R_b$ defines the Rydberg blockade radius (within which no two atoms can be simultaneously excited to the Rydberg state). Note that we are implicitly working in units where the lattice spacing is set to one. Since $V_{ij}$ falls off rapidly with the sixth power of the interatomic distance, we truncate the interactions beyond a  cutoff of third-nearest neighbors in our simulations, akin to Ref.~\onlinecite{Samajdar.2021}. We set $\Omega$\,$=$\,$1$ and scan the parameters $\delta$ and $R_b$ to explore the phase diagram, paying particular attention to the previously identified correlated region between the solid phases.

To solve the model in Eq.~\eqref{eq:eq1} in an unbiased manner, we modify and employ several stochastic series expansion (SSE) QMC schemes~\cite{Sandvik2003,ZY2019sweeping,ZY2020improved,ZY2020,ZY2021mixed,da2021phase,yan2022triangular,merali2021stochastic} to deal with such Rydberg arrays. By monitoring the behavior of various physical observables, e.g., correlation functions and structure factors, we map out the detailed phase diagram in Fig.~\ref{Fig1}. 
Our simulations are performed on the kagome lattice with periodic boundary conditions and system sizes $N=3L^2$ for linear dimensions $L=6,8,12$, while setting the inverse temperature $\beta=L$ to scale to the ground state. Besides the conventional observables employed in recent QMC work~\cite{yan2022triangular}, here, to reveal the intricate nature of the glassy phase and its transitions or crossovers to the neighboring phases, we employ different annealing, quench, and parallel tempering schemes~\cite{hukushima1996exchange,Brooke1999,Brooke2001,Sipser00,Preda2001,melko2007simulations,Das2008,yan2021sweeping,PhysRevB.63.184501,mitra2018quantum}. Scanning along  parameter paths in the phase diagram, we compute the Edwards-Anderson order parameter to detect the spin-glass behavior~\cite{Edwards1975,richards1984spin,georges2001quantum,binder1986spin}. More details about these schemes can be found in the Supplemental Material (SM)~\cite{suppl}.

\begin{figure}[b]
\centering
\includegraphics[width=\columnwidth]{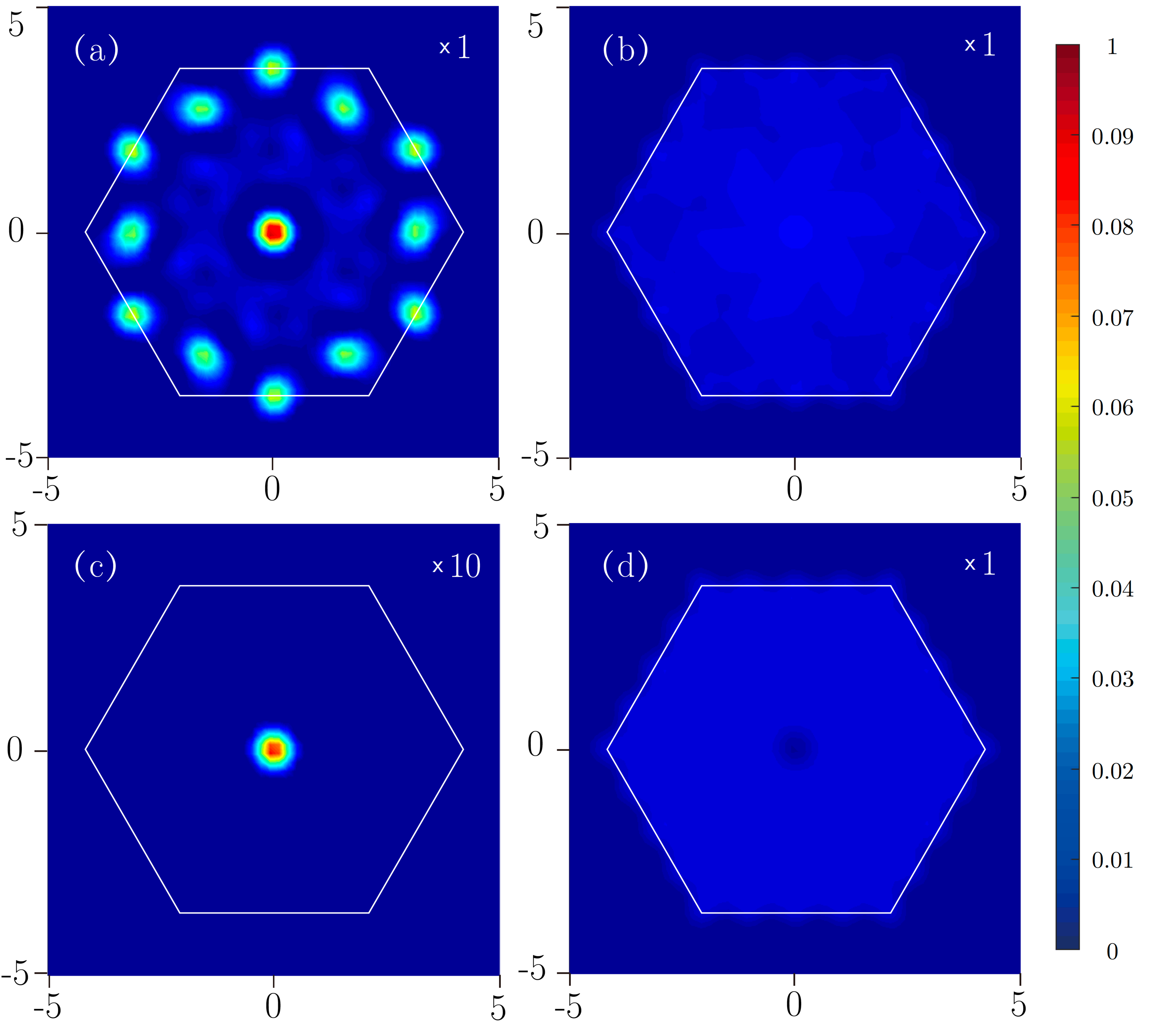}
\caption{Equal-time dimer structure factors $S(\mathbf{k},\tau$\,$=$\,$0)$ in the Brillouin zone for the (a) $1/6$ staggered ($R_b=2.3$), (b) glass ($R_b=2.1$), (c) nematic ($R_b=1.9$) and (d) PM ($R_b=1.05$) phases at $\delta/\Omega=3.3$ (vertical cut in the phase diagram of Fig.~\ref{Fig1}). The data shown here is simulated for $\beta$\,$=$\,$L$\,$=$\,$12$. The number in the upper-right corner shows the enlargement factor of the color bar; e.g., $\times 10$ changes the color bar from $[0, 0.1]$ to $[0, 1]$.
}
\label{Fig2}
\end{figure}

\begin{figure*}[t]
\centering
\includegraphics[width=\linewidth]{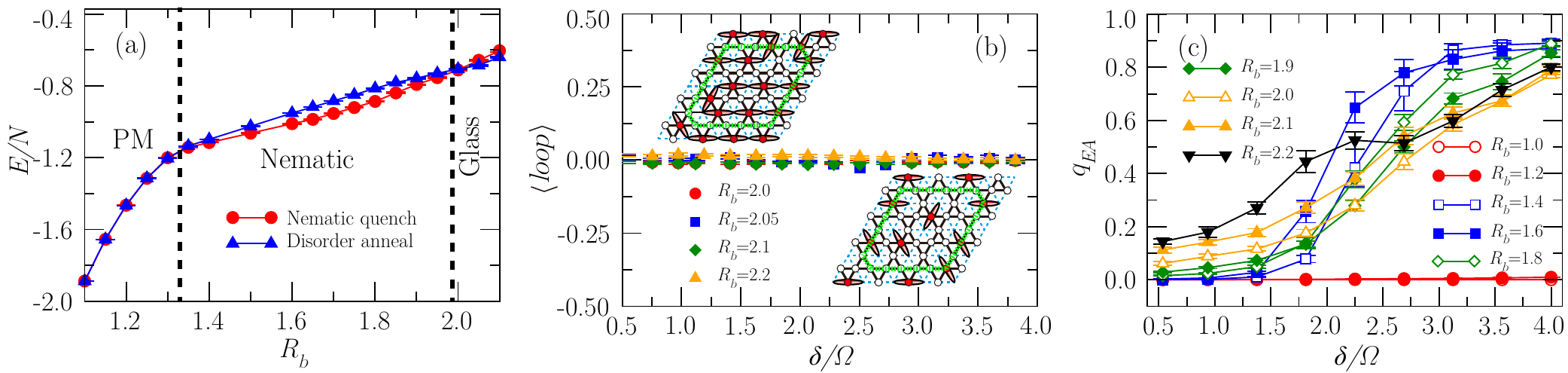}
\caption{(a) Energy per site plotted along the cut with $\delta/\Omega=3.3$ in Fig.~\ref{Fig1} using different initial states for the QMC simulations, showing that the nematic--glass phase transition is  first-order while the nematic--PM one is also weakly first-order. (b) The loop order parameter (green dashed string), which can be used to distinguish the even/odd $\mathbb{Z}_2$ QSL ($\langle loop \rangle\ne 0$) from the trivial disordered phase without topological order ($\langle loop \rangle =0$), is always close to zero, suggesting that the phase is not a pure even or odd $\mathbb{Z}_2$ QSL. (c) The Edwards-Anderson order parameter increases sharply upon going from the PM to the glass phase as $\delta/\Omega$ increases at different $R_b$.}
    \label{Fig3}
\end{figure*}

\textit{Phase diagram.---}The phase diagram thus obtained  is illustrated in Fig.~\ref{Fig1}. When $\delta/\Omega$ is small but positive, we observe a disordered paramagnetic (PM) phase. Once $\delta/\Omega$ is tuned to larger values, we find two symmetry-breaking VBS phases, in agreement with previous DMRG results~\cite{Samajdar.2021} but with slightly shifted phase boundaries. These solid phases, termed  nematic and staggered, correspond to (approximately) $1/3$ and $1/6$ filling of Rydberg excitations, respectively. The schematic plots of these crystalline phases are sketched in the insets of Fig.~\ref{Fig1}, both on the direct kagome lattice in the Rydberg basis and on the medial triangular lattice in the dimer basis. The exact manner in which they connect to each other, in  the thermodynamic limit, is an interesting open question, with possibilities including topologically ordered even or odd quantum spin liquids (QSLs), an intervening trivial disordered phase, or some new emergent intermediate phase \cite{Samajdar.2021,yan2022triangular}.

Interestingly, we discover that a glassy disordered phase---which can be distinguished from the PM by the magnitude of the Edwards-Anderson order parameter--- exists in the central region between the two VBSs. The phase boundaries between this region and proximate phases are determined by examining various parameter points and paths scanning through the phase diagram, as denoted by the red stars and dashed lines in Fig.~\ref{Fig1}, and addressed in detail below.

To characterize the variety of phases, we first compute the equal-time $(\tau$\,$=$\,$0)$ structure factor (see Fig.~\ref{Fig2}) as
\begin{equation}
S(\mathbf{k},\tau)=\frac{1}{N}
\hspace*{-0.15cm}
\sum_{\substack{i,j \\ \alpha=1,2,3}}^{L^3}
\hspace*{-0.15cm}
e^{i\mathbf{k}\cdot \mathbf{r}_{ij}} \bigg(\langle n^{}_{i,\alpha}(\tau)n^{}_{j,\alpha}(0)\rangle - \langle n^{}_{i,\alpha}\rangle \langle n^{}_{j,\alpha}\rangle \bigg),
\end{equation}
where $n_i$ is the density operator on site $i$ and $\alpha$ stands for the three sublattices of the kagome lattice, at four representative parameter points corresponding to the four distinct phases in the phase diagram. Figures~\ref{Fig2} (b) and (d) show $S(\mathbf{k},0)$ inside the glass and PM phases, respectively. In the hexagonal Brillouin zone, we observe that there are no peaks signifying long-range order but only broad profiles associated with different short-range density correlation patterns in real space. In contrast, Figs.~\ref{Fig2} (a) and (c) present the structure factors inside the staggered and nematic phases, respectively, where one now clearly sees the Bragg peaks at the relevant ordering wavevectors.

\textit{Quantum phase transitions.---}Having established the lack of long-range density correlations in both the PM and glass phases, we move on to study the associated quantum phase transitions~\cite{sachdev2011quantum}. Since the glass phase is expected to have many degenerate energy minima and very long autocorrelation times (which render the QMC simulation difficult), special care needs to be taken in determining its phase boundaries. Our results in this regard are summarized in Fig.~\ref{Fig3}, which shows the data along several parameter scans in the phase diagram (dashed lines in Fig.~\ref{Fig1}).

First, in Fig.~\ref{Fig3}(a), we illustrate the energy density along the line $\delta/\Omega=3.3$, computed with different initial states and annealing/quench schemes, to find the phase transition between the nematic and PM phases. The red curve shows the energy density simulated from random initial configurations with thermal annealing (by decreasing the temperature  slowly)~\cite{Brooke1999,Brooke2001}. On the other hand, the data plotted in blue is simulated from nematic configurations by quenching (i.e., starting at a very low temperature). Deep in the nematic phase, the two energy lines are clearly distinct. The difference between the two becomes small on progressing towards the transition point, where the two energy lines cross and then split weakly in the PM phase. Thus, the phase transition between the PM and nematic phases, which belongs to the $(2$\,$+$\,$1)$D three-state Potts universality class~\cite{Samajdar.2021}, is seen to be weakly first-order in consistency with prior findings~\cite{janke1997three}.
This first-order phase transition can also be detected from the order parameter of the nematic phase, as detailed in the SM~\cite{suppl}. By the same logic, Fig.~\ref{Fig3}(a) also conveys that the transition between the glass and the nematic phase, which occurs at $ R_b$\,$\sim$\,$2$ in Fig.~\ref{Fig1}, is first-order as well. Scanning the energy density along the line $R_b=1.9$, as shown in the SM~\cite{suppl}, similarly manifests a first-order phase transition.

We now study the central disordered region between the two VBS phases, considering, in particular, the possible even and odd QSLs, or PM phases that emerge in an approximate quantum dimer model~\cite{yan2022triangular}. To this end, we define a nonlocal loop operator~\cite{Semeghini21}---schematically shown by the green dashed loop in the inset of Fig.~\ref{Fig3}(b)---as $\langle loop \rangle= \langle (-1)^{\#~\mathrm{cut~dimers}} \rangle$, which measures the parity of the number of dimers intersected along a rhomboid with odd linear size on the medial dimer lattice. This operator can be used to distinguish the two QSLs and the PM phase~\cite{Semeghini21,yan2022triangular}. In an odd (even) $\mathbb{Z}_2$ QSL without spinon excitations, the value of $\langle loop \rangle$ is  pinned to $-1$ ($+1$) because of the exact constraint requiring one (two) dimer(s) per site of the triangular lattice; this operator continues to be well-defined for a small density of matter fields~\cite{IGT}. From Fig.~\ref{Fig3}(b), we see that $\langle loop \rangle$ remains close to zero in the central correlated region, indicating that the ground state is not a pure even or odd $\mathbb{Z}_2$ QSL. As shown below, we further find that this region is also not a trivial PM phase, but, perhaps surprisingly for a homogeneous Hamiltonian, an emergent glass phase. This finding also underscores that the low-energy effective theory---a triangular lattice quantum dimer model with variable dimer density~\cite{yan2022triangular}---used to describe the physics proximate to the VBS phases departs from the realistic Rydberg Hamiltonian in this part of the phase diagram, where the Rydberg excitation density differs significantly from $1/6$ or $1/3$ (corresponding to the limit of one or two dimers per site, respectively). Moreover, the snapshots of the sampled configurations drawn in the SM~\cite{suppl} also demonstrate that apart from the Rydberg blockade, all local constraints (associated with the $\mathbb{Z}_2$ topological order) are relaxed in this glass region.

\begin{figure}[tb]
\centering
\includegraphics[width=\linewidth]{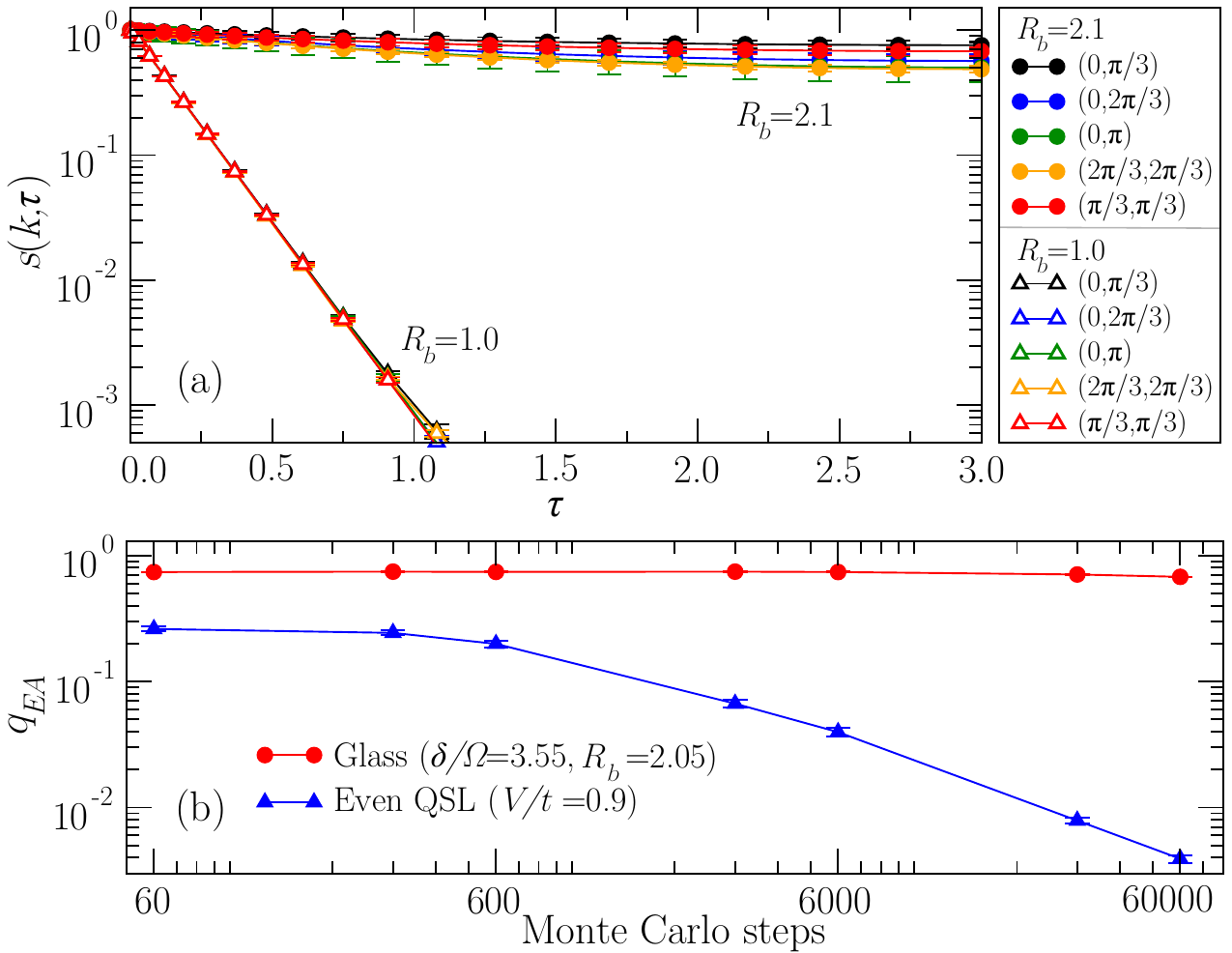}
\caption{(a) Dynamical correlations $S(\mathbf{k},\tau)$ deep in the glass  ($\delta/\Omega$\,$=$\,$3.3$,\,$R_b$\,$=$\,$2.1$) and PM ($\delta/\Omega$\,$=$\,$3.3$,\,$R_b$\,$=$\,$1$) phases. The correlations decay quickly with similar slopes in the PM phase, indicating a large gap and a flat band. Meanwhile, the correlations in the glass phase decay very slowly, which suggests that there are many nearly degenerate local minima in the energy landscape. (b) The Edwards-Anderson order parameter $q^{}_{\textsc{ea}}$ also decays much more slowly with the increasing number of Monte Carlo steps for the glass than for the QSL. The data in the glass phase is computed for the model~\eqref{eq:eq1} at $\delta/\Omega=3.55$, $R_b=2.05$, while the QSL's data is obtained from the triangular-lattice quantum loop model at $V/t=0.9$. 
All the data are simulated at $\beta=6$ on a $L=6$ lattice.}
\label{Fig4}
\end{figure}

To differentiate between the PM and glass phases, we utilize the Edwards-Anderson order parameter~\cite{Edwards1975,richards1984spin,binder1986spin,georges2001quantum,king2022quantum}
$
q^{}_{\textsc{ea}}=\sum_{i=1}^N\langle n^{}_i-\rho\rangle^2/[N\rho(1-\rho)]
$,
where $n_i \equiv\left|r\right>_i\left<r\right|$, $\rho$ is the average density defined as $\rho \equiv \sum_{i=1}^N\langle n_i\rangle/N$, and $\langle \cdots \rangle$ indicates a statistical average over Monte Carlo snapshots~\cite{PhysRevLett.103.215302,angelone2016superglass}. The spin-glass order is characterized  by the breaking of translational invariance and the Edwards-Anderson order parameter $q^{}_{\textsc{ea}}$\,$\in$\,$[0, 1]$, which captures the on-site deviation from the average density, is a measure of the glassy behavior. The large magnitude of $q^{}_{\textsc{ea}}$ shown in Fig.~\ref{Fig3}(c) demonstrates the emergent glassy nature of the disordered region amid the VBS phases. While $q^{}_{\textsc{ea}}$ clearly tells the PM and glass phases apart, whether the two are separated by a phase transition or by a crossover is an interesting open question. It is also noteworthy that this Edwards-Anderson order parameter decays extremely slow with increasing the number of Monte Carlo steps [see Fig.~\ref{Fig4}(b)]; this is another signature of the slow dynamics in the glass phase~\cite{keren1996probing}, which is consistent with the small and nearly degenerate gaps that we will now establish. 

\textit{Glassy dynamics.---}One of the hallmarks of a quantum glass is its ability to support nearly gapless excitations at all momenta due to the existence of exponentially many local minima in its energy landscape~\cite{malinovsky1991log,keren1996probing,rozenberg1998dynamics,ma2018spin-glass}. Meanwhile, the PM phase is obviously gapped and dispersionless with a short correlation length. In this section, we focus on the measurement of imaginary-time correlations $S(\mathbf{k},\tau)$ at different momenta, deep inside the PM ($\delta/\Omega=3.3$, $R_b$\,$=$\,$1.0$) and glass ($\delta/\Omega$\,$=$\,$3.3$, $R_b$\,$=$\,$2.1$) phases, as marked by the red stars (d) and (b) in Fig.~\ref{Fig1}, respectively. Figure~\ref{Fig4}(a) highlights a striking distinction in $S(\mathbf{k},\tau)$ between these two regions. The upper ones, which are measured inside the glass phase, decay slowly, indicating that their gaps are all very small and almost equal. On the other hand, the lower ones measuring the correlations in PM phase decay much more quickly with similar slopes. These two kinds of correlations are consistent with the respective features of the glass phase, which possesses nearly gapless excitations, as well as the PM phase, which hosts gapped flat bands.

Lastly, we also compare the behavior of the Edwards-Anderson order parameter $q^{}_{\textsc{ea}}$, as a function of the number of Monte Carlo steps, inside the glass phase to that in an even QSL~\cite{ZY2022loop}. 
Since the representative parameter point chosen for the spin glass ($\delta/\Omega$\,$=$\,$3.55$,\,$R_b$\,$=$\,$2.05$) lies close to the boundary of the nematic phase (which, recall, is a VBS with two dimers per site), the low-energy effective model describing the QSL is taken to be the triangular-lattice quantum loop model close to the Rokhsar-Kivelson point~\cite{yan2022triangular,ZY2022loop} ($V/t$\,$=$\,$0.9$, where $V$ is the interaction between parallel dimers and $-t$ is the dimer resonance energy). As shown in Fig.~\ref{Fig4}(b), in the QSL phase, $q^{}_{\textsc{ea}}$ is not only an order-of-magnitude smaller but also decays much faster than in the glass. The Monte Carlo \textit{dynamics} of $q^{}_{\textsc{ea}}$ thus complement the \textit{static} results of Fig.~\ref{Fig3}(b) and together, highlight the distinction between the QSL and glass phases.

\textit{Discussion.---}In this work, we investigated a realistic Rydberg Hamiltonian on the kagome lattice with large-scale QMC simulations and uncover---besides two VBSs and a PM phase---an emergent glass phase in the phase diagram. The origin of glassiness is likely due to the kinetic constraints associated with the Rydberg blockade that forbid several hopping processes~\cite{ritort2003glassy,lesanovsky2013kinetic,knap2019}. Such a glass phase constitutes a new addition to the growing list of correlated quantum many-body states that can be studied on current-generation Rydberg platforms. Through detailed QMC analyses, we explore the subtle behavior of this glass phase with its intricate degenerate energy landscapes, establish its unique static and dynamic fingerprints, and compare and contrast its properties to those of competing QSL candidates. However, we note that the observation of a glassy \textit{ground state} does not necessarily rule out the dynamical preparation of QSL states, which may be obtained as a macroscopic superposition of dimer configurations during quasiadiabatic sweeps in experiments~\cite{Semeghini21,hannes_dynamical,zhai}.

Experimentally, the spin glass phase can be detected by preparing a (deterministic) far-from-equilibrium initial state, quenching to the glassy region, measuring snapshots in the occupation basis, and repeating the protocol, but stopping at different  points in the temporal evolution each time. This would allow one to observe the anomalously slow relaxation dynamics, as recently demonstrated in experiments studying a disordered XXZ model on a Rydberg quantum simulator~\cite{signoles2021glassy}. 
It is also possible to use Bragg spectroscopy to measure the dynamics of the glass phase in cold-atom systems held in optical lattices~\cite{senaratneSpin2022}.
Finally, even though the Edwards-Anderson order parameter can be challenging to measure in quantum simulators (since it requires knowledge of two-time correlation functions), one can probe an efficient proxy for $q^{}_{\textsc{ea}}$ by measuring the eigenstate spin-glass order parameter~\cite{heyl2019} constructed from two-site reduced density matrices, which can be can be accessed by a variety of methods including quantum state tomography~\cite{huang2020predicting}.

\begin{acknowledgments}
\textit{Acknowledgments.---}Z.Y. and Z.Y.M. acknowledge useful discussions with Adriano Angelone, Xue-Feng Zhang and Zheng Zhou.
R.S. and S.S. thank their coauthors in earlier collaborations \cite{Samajdar.2021,Semeghini21} for helpful discussions. R.S. is supported by the Princeton Quantum Initiative Fellowship. S.S. is supported by the U.S. Department of Energy under Grant DE-SC0019030.  ZY and ZYM acknowledge support from the RGC of Hong Kong SAR of China (Project Nos. 17301420, 17301721, AoE/P-701/20, 17309822,  HKU C7037-22G), the ANR/RGC Joint Research Scheme sponsored by Research Grants Council of Hong Kong SAR of China and French National Research Agency(Project No. A\_HKU703/22), the K. C. Wong Education Foundation (Grant No.~GJTD-2020-01), and the Seed Funding ``Quantum-Inspired explainable-AI" at
the HKU-TCL Joint Research Centre for Artificial Intelligence. Y.C.W. acknowledges  support from Zhejiang Provincial Natural Science Foundation of China (Grant Nos. LZ23A040003) and the  Beihang Hangzhou Innovation Institute Yuhang. We thank Beijng PARATERA Tech CO.,Ltd., the High-performance Computing Centre of Beihang Hangzhou Innovation Institute Yuhang, HPC2021 system under the Information Technology Services, and the Blackbody HPC system at the Department of Physics, University of Hong Kong for providing computational resources that have contributed to the research results in this paper.
\end{acknowledgments}

\bibliographystyle{apsrev4-2_custom}
\bibliography{refs.bib}

\begin{thebibliography}{80}%
\makeatletter
\providecommand \@ifxundefined [1]{%
 \@ifx{#1\undefined}
}%
\providecommand \@ifnum [1]{%
 \ifnum #1\expandafter \@firstoftwo
 \else \expandafter \@secondoftwo
 \fi
}%
\providecommand \@ifx [1]{%
 \ifx #1\expandafter \@firstoftwo
 \else \expandafter \@secondoftwo
 \fi
}%
\providecommand \natexlab [1]{#1}%
\providecommand \enquote  [1]{``#1''}%
\providecommand \bibnamefont  [1]{#1}%
\providecommand \bibfnamefont [1]{#1}%
\providecommand \citenamefont [1]{#1}%
\providecommand \href@noop [0]{\@secondoftwo}%
\providecommand \href [0]{\begingroup \@sanitize@url \@href}%
\providecommand \@href[1]{\@@startlink{#1}\@@href}%
\providecommand \@@href[1]{\endgroup#1\@@endlink}%
\providecommand \@sanitize@url [0]{\catcode `\\12\catcode `\$12\catcode
  `\&12\catcode `\#12\catcode `\^12\catcode `\_12\catcode `\%12\relax}%
\providecommand \@@startlink[1]{}%
\providecommand \@@endlink[0]{}%
\providecommand \url  [0]{\begingroup\@sanitize@url \@url }%
\providecommand \@url [1]{\endgroup\@href {#1}{\urlprefix }}%
\providecommand \urlprefix  [0]{URL }%
\providecommand \Eprint [0]{\href }%
\providecommand \doibase [0]{https://doi.org/}%
\providecommand \selectlanguage [0]{\@gobble}%
\providecommand \bibinfo  [0]{\@secondoftwo}%
\providecommand \bibfield  [0]{\@secondoftwo}%
\providecommand \translation [1]{[#1]}%
\providecommand \BibitemOpen [0]{}%
\providecommand \bibitemStop [0]{}%
\providecommand \bibitemNoStop [0]{.\EOS\space}%
\providecommand \EOS [0]{\spacefactor3000\relax}%
\providecommand \BibitemShut  [1]{\csname bibitem#1\endcsname}%
\let\auto@bib@innerbib\@empty
\bibitem [{\citenamefont {Endres}\ \emph {et~al.}(2016)\citenamefont {Endres},
  \citenamefont {Bernien}, \citenamefont {Keesling}, \citenamefont {Levine},
  \citenamefont {Anschuetz}, \citenamefont {Krajenbrink}, \citenamefont
  {Senko}, \citenamefont {Vuleti\'c}, \citenamefont {Greiner},\ and\
  \citenamefont {Lukin}}]{endres2016atom}%
  \BibitemOpen
  \bibfield  {author} {\bibinfo {author} {\bibfnamefont {M.}~\bibnamefont
  {Endres}}, \bibinfo {author} {\bibfnamefont {H.}~\bibnamefont {Bernien}},
  \bibinfo {author} {\bibfnamefont {A.}~\bibnamefont {Keesling}}, \bibinfo
  {author} {\bibfnamefont {H.}~\bibnamefont {Levine}}, \bibinfo {author}
  {\bibfnamefont {E.~R.}\ \bibnamefont {Anschuetz}}, \bibinfo {author}
  {\bibfnamefont {A.}~\bibnamefont {Krajenbrink}}, \bibinfo {author}
  {\bibfnamefont {C.}~\bibnamefont {Senko}}, \bibinfo {author} {\bibfnamefont
  {V.}~\bibnamefont {Vuleti\'c}}, \bibinfo {author} {\bibfnamefont
  {M.}~\bibnamefont {Greiner}},\ and\ \bibinfo {author} {\bibfnamefont {M.~D.}\
  \bibnamefont {Lukin}},\ }\bibfield  {title} {\bibinfo {title} {Atom-by-atom
  assembly of defect-free one-dimensional cold atom arrays},\ }\href
  {https://doi.org/10.1126/science.aah3752} {\bibfield  {journal} {\bibinfo
  {journal} {Science}\ }\textbf {\bibinfo {volume} {354}},\ \bibinfo {pages}
  {1024} (\bibinfo {year} {2016})}\BibitemShut {NoStop}%
\bibitem [{\citenamefont {Bernien}\ \emph {et~al.}(2017)\citenamefont
  {Bernien}, \citenamefont {Schwartz}, \citenamefont {Keesling}, \citenamefont
  {Levine}, \citenamefont {Omran}, \citenamefont {Pichler}, \citenamefont
  {Choi}, \citenamefont {Zibrov}, \citenamefont {Endres}, \citenamefont
  {Greiner}, \citenamefont {Vuleti{\'c}},\ and\ \citenamefont
  {Lukin}}]{bernien2017probing}%
  \BibitemOpen
  \bibfield  {author} {\bibinfo {author} {\bibfnamefont {H.}~\bibnamefont
  {Bernien}}, \bibinfo {author} {\bibfnamefont {S.}~\bibnamefont {Schwartz}},
  \bibinfo {author} {\bibfnamefont {A.}~\bibnamefont {Keesling}}, \bibinfo
  {author} {\bibfnamefont {H.}~\bibnamefont {Levine}}, \bibinfo {author}
  {\bibfnamefont {A.}~\bibnamefont {Omran}}, \bibinfo {author} {\bibfnamefont
  {H.}~\bibnamefont {Pichler}}, \bibinfo {author} {\bibfnamefont
  {S.}~\bibnamefont {Choi}}, \bibinfo {author} {\bibfnamefont {A.~S.}\
  \bibnamefont {Zibrov}}, \bibinfo {author} {\bibfnamefont {M.}~\bibnamefont
  {Endres}}, \bibinfo {author} {\bibfnamefont {M.}~\bibnamefont {Greiner}},
  \bibinfo {author} {\bibfnamefont {V.}~\bibnamefont {Vuleti{\'c}}},\ and\
  \bibinfo {author} {\bibfnamefont {M.~D.}\ \bibnamefont {Lukin}},\ }\bibfield
  {title} {\bibinfo {title} {Probing many-body dynamics on a 51-atom quantum
  simulator},\ }\href {https://doi.org/10.1038/nature24622} {\bibfield
  {journal} {\bibinfo  {journal} {Nature}\ }\textbf {\bibinfo {volume} {551}},\
  \bibinfo {pages} {579} (\bibinfo {year} {2017})}\BibitemShut {NoStop}%
\bibitem [{\citenamefont {Browaeys}\ and\ \citenamefont
  {Lahaye}(2020)}]{Browaeys.2020}%
  \BibitemOpen
  \bibfield  {author} {\bibinfo {author} {\bibfnamefont {A.}~\bibnamefont
  {Browaeys}}\ and\ \bibinfo {author} {\bibfnamefont {T.}~\bibnamefont
  {Lahaye}},\ }\bibfield  {title} {\bibinfo {title} {{Many-body physics with
  individually controlled Rydberg atoms}},\ }\href
  {https://doi.org/10.1038/s41567-019-0733-z} {\bibfield  {journal} {\bibinfo
  {journal} {Nat. Phys.}\ }\textbf {\bibinfo {volume} {16}},\ \bibinfo {pages}
  {132} (\bibinfo {year} {2020})}\BibitemShut {NoStop}%
\bibitem [{\citenamefont {Morgado}\ and\ \citenamefont
  {Whitlock}(2021)}]{morgado2021quantum}%
  \BibitemOpen
  \bibfield  {author} {\bibinfo {author} {\bibfnamefont {M.}~\bibnamefont
  {Morgado}}\ and\ \bibinfo {author} {\bibfnamefont {S.}~\bibnamefont
  {Whitlock}},\ }\bibfield  {title} {\bibinfo {title} {{Quantum simulation and
  computing with Rydberg-interacting qubits}},\ }\href
  {https://doi.org/10.1116/5.0036562} {\bibfield  {journal} {\bibinfo
  {journal} {AVS Quantum Sci.}\ }\textbf {\bibinfo {volume} {3}},\ \bibinfo
  {pages} {023501} (\bibinfo {year} {2021})}\BibitemShut {NoStop}%
\bibitem [{\citenamefont {Bluvstein}\ \emph {et~al.}(2022)\citenamefont
  {Bluvstein}, \citenamefont {Levine}, \citenamefont {Semeghini}, \citenamefont
  {Wang}, \citenamefont {Ebadi}, \citenamefont {Kalinowski}, \citenamefont
  {Keesling}, \citenamefont {Maskara}, \citenamefont {Pichler}, \citenamefont
  {Greiner}, \citenamefont {Vuletic},\ and\ \citenamefont
  {Lukin}}]{bluvstein2022quantum}%
  \BibitemOpen
  \bibfield  {author} {\bibinfo {author} {\bibfnamefont {D.}~\bibnamefont
  {Bluvstein}}, \bibinfo {author} {\bibfnamefont {H.}~\bibnamefont {Levine}},
  \bibinfo {author} {\bibfnamefont {G.}~\bibnamefont {Semeghini}}, \bibinfo
  {author} {\bibfnamefont {T.~T.}\ \bibnamefont {Wang}}, \bibinfo {author}
  {\bibfnamefont {S.}~\bibnamefont {Ebadi}}, \bibinfo {author} {\bibfnamefont
  {M.}~\bibnamefont {Kalinowski}}, \bibinfo {author} {\bibfnamefont
  {A.}~\bibnamefont {Keesling}}, \bibinfo {author} {\bibfnamefont
  {N.}~\bibnamefont {Maskara}}, \bibinfo {author} {\bibfnamefont
  {H.}~\bibnamefont {Pichler}}, \bibinfo {author} {\bibfnamefont
  {M.}~\bibnamefont {Greiner}}, \bibinfo {author} {\bibfnamefont
  {V.}~\bibnamefont {Vuletic}},\ and\ \bibinfo {author} {\bibfnamefont {M.~D.}\
  \bibnamefont {Lukin}},\ }\bibfield  {title} {\bibinfo {title} {A quantum
  processor based on coherent transport of entangled atom arrays},\ }\href
  {https://doi.org/10.48550/arXiv.2112.03923} {\bibfield  {journal} {\bibinfo
  {journal} {Nature}\ }\textbf {\bibinfo {volume} {604}},\ \bibinfo {pages}
  {451} (\bibinfo {year} {2022})}\BibitemShut {NoStop}%
\bibitem [{\citenamefont {de~L{\'e}s{\'e}leuc}\ \emph
  {et~al.}(2019)\citenamefont {de~L{\'e}s{\'e}leuc}, \citenamefont {Lienhard},
  \citenamefont {Scholl}, \citenamefont {Barredo}, \citenamefont {Weber},
  \citenamefont {Lang}, \citenamefont {B{\"u}chler}, \citenamefont {Lahaye},\
  and\ \citenamefont {Browaeys}}]{de2019observation}%
  \BibitemOpen
  \bibfield  {author} {\bibinfo {author} {\bibfnamefont {S.}~\bibnamefont
  {de~L{\'e}s{\'e}leuc}}, \bibinfo {author} {\bibfnamefont {V.}~\bibnamefont
  {Lienhard}}, \bibinfo {author} {\bibfnamefont {P.}~\bibnamefont {Scholl}},
  \bibinfo {author} {\bibfnamefont {D.}~\bibnamefont {Barredo}}, \bibinfo
  {author} {\bibfnamefont {S.}~\bibnamefont {Weber}}, \bibinfo {author}
  {\bibfnamefont {N.}~\bibnamefont {Lang}}, \bibinfo {author} {\bibfnamefont
  {H.~P.}\ \bibnamefont {B{\"u}chler}}, \bibinfo {author} {\bibfnamefont
  {T.}~\bibnamefont {Lahaye}},\ and\ \bibinfo {author} {\bibfnamefont
  {A.}~\bibnamefont {Browaeys}},\ }\bibfield  {title} {\bibinfo {title}
  {{Observation of a symmetry-protected topological phase of interacting bosons
  with Rydberg atoms}},\ }\href {https://doi.org/10.1126/science.aav9105}
  {\bibfield  {journal} {\bibinfo  {journal} {Science}\ }\textbf {\bibinfo
  {volume} {365}},\ \bibinfo {pages} {775} (\bibinfo {year}
  {2019})}\BibitemShut {NoStop}%
\bibitem [{\citenamefont {Ebadi}\ \emph {et~al.}(2021)\citenamefont {Ebadi},
  \citenamefont {Wang}, \citenamefont {Levine}, \citenamefont {Keesling},
  \citenamefont {Semeghini}, \citenamefont {Omran}, \citenamefont {Bluvstein},
  \citenamefont {Samajdar}, \citenamefont {Pichler}, \citenamefont {Ho},
  \citenamefont {Choi}, \citenamefont {Sachdev}, \citenamefont {Greiner},
  \citenamefont {Vuleti\'c},\ and\ \citenamefont {Lukin}}]{Ebadi.2021}%
  \BibitemOpen
  \bibfield  {author} {\bibinfo {author} {\bibfnamefont {S.}~\bibnamefont
  {Ebadi}}, \bibinfo {author} {\bibfnamefont {T.~T.}\ \bibnamefont {Wang}},
  \bibinfo {author} {\bibfnamefont {H.}~\bibnamefont {Levine}}, \bibinfo
  {author} {\bibfnamefont {A.}~\bibnamefont {Keesling}}, \bibinfo {author}
  {\bibfnamefont {G.}~\bibnamefont {Semeghini}}, \bibinfo {author}
  {\bibfnamefont {A.}~\bibnamefont {Omran}}, \bibinfo {author} {\bibfnamefont
  {D.}~\bibnamefont {Bluvstein}}, \bibinfo {author} {\bibfnamefont
  {R.}~\bibnamefont {Samajdar}}, \bibinfo {author} {\bibfnamefont
  {H.}~\bibnamefont {Pichler}}, \bibinfo {author} {\bibfnamefont {W.~W.}\
  \bibnamefont {Ho}}, \bibinfo {author} {\bibfnamefont {S.}~\bibnamefont
  {Choi}}, \bibinfo {author} {\bibfnamefont {S.}~\bibnamefont {Sachdev}},
  \bibinfo {author} {\bibfnamefont {M.}~\bibnamefont {Greiner}}, \bibinfo
  {author} {\bibfnamefont {V.}~\bibnamefont {Vuleti\'c}},\ and\ \bibinfo
  {author} {\bibfnamefont {M.~D.}\ \bibnamefont {Lukin}},\ }\bibfield  {title}
  {\bibinfo {title} {{Quantum phases of matter on a 256-atom programmable
  quantum simulator}},\ }\href {https://doi.org/10.1038/s41586-021-03582-4}
  {\bibfield  {journal} {\bibinfo  {journal} {Nature}\ }\textbf {\bibinfo
  {volume} {595}},\ \bibinfo {pages} {227} (\bibinfo {year}
  {2021})}\BibitemShut {NoStop}%
\bibitem [{\citenamefont {Scholl}\ \emph {et~al.}(2021)\citenamefont {Scholl},
  \citenamefont {Schuler}, \citenamefont {Williams}, \citenamefont
  {Eberharter}, \citenamefont {Barredo}, \citenamefont {Schymik}, \citenamefont
  {Lienhard}, \citenamefont {Henry}, \citenamefont {Lang}, \citenamefont
  {Lahaye}, \citenamefont {L\"auchli},\ and\ \citenamefont
  {Browaeys}}]{scholl2021quantum}%
  \BibitemOpen
  \bibfield  {author} {\bibinfo {author} {\bibfnamefont {P.}~\bibnamefont
  {Scholl}}, \bibinfo {author} {\bibfnamefont {M.}~\bibnamefont {Schuler}},
  \bibinfo {author} {\bibfnamefont {H.~J.}\ \bibnamefont {Williams}}, \bibinfo
  {author} {\bibfnamefont {A.~A.}\ \bibnamefont {Eberharter}}, \bibinfo
  {author} {\bibfnamefont {D.}~\bibnamefont {Barredo}}, \bibinfo {author}
  {\bibfnamefont {K.-N.}\ \bibnamefont {Schymik}}, \bibinfo {author}
  {\bibfnamefont {V.}~\bibnamefont {Lienhard}}, \bibinfo {author}
  {\bibfnamefont {L.-P.}\ \bibnamefont {Henry}}, \bibinfo {author}
  {\bibfnamefont {T.~C.}\ \bibnamefont {Lang}}, \bibinfo {author}
  {\bibfnamefont {T.}~\bibnamefont {Lahaye}}, \bibinfo {author} {\bibfnamefont
  {A.~M.}\ \bibnamefont {L\"auchli}},\ and\ \bibinfo {author} {\bibfnamefont
  {A.}~\bibnamefont {Browaeys}},\ }\bibfield  {title} {\bibinfo {title}
  {{Quantum simulation of 2D antiferromagnets with hundreds of Rydberg
  atoms}},\ }\href {https://doi.org/10.1038/s41586-021-03585-1} {\bibfield
  {journal} {\bibinfo  {journal} {Nature}\ }\textbf {\bibinfo {volume} {595}},\
  \bibinfo {pages} {233} (\bibinfo {year} {2021})}\BibitemShut {NoStop}%
\bibitem [{\citenamefont {{Semeghini}}\ \emph {et~al.}(2021)\citenamefont
  {{Semeghini}}, \citenamefont {{Levine}}, \citenamefont {{Keesling}},
  \citenamefont {{Ebadi}}, \citenamefont {{Wang}}, \citenamefont {{Bluvstein}},
  \citenamefont {{Verresen}}, \citenamefont {{Pichler}}, \citenamefont
  {{Kalinowski}}, \citenamefont {{Samajdar}}, \citenamefont {{Omran}},
  \citenamefont {{Sachdev}}, \citenamefont {{Vishwanath}}, \citenamefont
  {{Greiner}}, \citenamefont {{Vuleti{\'c}}},\ and\ \citenamefont
  {{Lukin}}}]{Semeghini21}%
  \BibitemOpen
  \bibfield  {author} {\bibinfo {author} {\bibfnamefont {G.}~\bibnamefont
  {{Semeghini}}}, \bibinfo {author} {\bibfnamefont {H.}~\bibnamefont
  {{Levine}}}, \bibinfo {author} {\bibfnamefont {A.}~\bibnamefont
  {{Keesling}}}, \bibinfo {author} {\bibfnamefont {S.}~\bibnamefont {{Ebadi}}},
  \bibinfo {author} {\bibfnamefont {T.~T.}\ \bibnamefont {{Wang}}}, \bibinfo
  {author} {\bibfnamefont {D.}~\bibnamefont {{Bluvstein}}}, \bibinfo {author}
  {\bibfnamefont {R.}~\bibnamefont {{Verresen}}}, \bibinfo {author}
  {\bibfnamefont {H.}~\bibnamefont {{Pichler}}}, \bibinfo {author}
  {\bibfnamefont {M.}~\bibnamefont {{Kalinowski}}}, \bibinfo {author}
  {\bibfnamefont {R.}~\bibnamefont {{Samajdar}}}, \bibinfo {author}
  {\bibfnamefont {A.}~\bibnamefont {{Omran}}}, \bibinfo {author} {\bibfnamefont
  {S.}~\bibnamefont {{Sachdev}}}, \bibinfo {author} {\bibfnamefont
  {A.}~\bibnamefont {{Vishwanath}}}, \bibinfo {author} {\bibfnamefont
  {M.}~\bibnamefont {{Greiner}}}, \bibinfo {author} {\bibfnamefont
  {V.}~\bibnamefont {{Vuleti{\'c}}}},\ and\ \bibinfo {author} {\bibfnamefont
  {M.~D.}\ \bibnamefont {{Lukin}}},\ }\bibfield  {title} {\bibinfo {title}
  {{Probing topological spin liquids on a programmable quantum simulator}},\
  }\href {https://doi.org/10.1126/science.abi8794} {\bibfield  {journal}
  {\bibinfo  {journal} {Science}\ }\textbf {\bibinfo {volume} {374}},\ \bibinfo
  {pages} {1242} (\bibinfo {year} {2021})}\BibitemShut {NoStop}%
\bibitem [{\citenamefont {Turner}\ \emph {et~al.}(2018)\citenamefont {Turner},
  \citenamefont {Michailidis}, \citenamefont {Abanin}, \citenamefont {Serbyn},\
  and\ \citenamefont {Papi{\'c}}}]{turner2018weak}%
  \BibitemOpen
  \bibfield  {author} {\bibinfo {author} {\bibfnamefont {C.~J.}\ \bibnamefont
  {Turner}}, \bibinfo {author} {\bibfnamefont {A.~A.}\ \bibnamefont
  {Michailidis}}, \bibinfo {author} {\bibfnamefont {D.~A.}\ \bibnamefont
  {Abanin}}, \bibinfo {author} {\bibfnamefont {M.}~\bibnamefont {Serbyn}},\
  and\ \bibinfo {author} {\bibfnamefont {Z.}~\bibnamefont {Papi{\'c}}},\
  }\bibfield  {title} {\bibinfo {title} {Weak ergodicity breaking from quantum
  many-body scars},\ }\href {https://doi.org/10.1038/s41567-018-0137-5}
  {\bibfield  {journal} {\bibinfo  {journal} {Nature Phys.}\ }\textbf {\bibinfo
  {volume} {14}},\ \bibinfo {pages} {745} (\bibinfo {year} {2018})}\BibitemShut
  {NoStop}%
\bibitem [{\citenamefont {Keesling}\ \emph {et~al.}(2019)\citenamefont
  {Keesling}, \citenamefont {Omran}, \citenamefont {Levine}, \citenamefont
  {Bernien}, \citenamefont {Pichler}, \citenamefont {Choi}, \citenamefont
  {Samajdar}, \citenamefont {Schwartz}, \citenamefont {Silvi}, \citenamefont
  {Sachdev}, \citenamefont {Zoller}, \citenamefont {Endres}, \citenamefont
  {Greiner}, \citenamefont {Vuleti{\'c}},\ and\ \citenamefont
  {Lukin}}]{keesling2019quantum}%
  \BibitemOpen
  \bibfield  {author} {\bibinfo {author} {\bibfnamefont {A.}~\bibnamefont
  {Keesling}}, \bibinfo {author} {\bibfnamefont {A.}~\bibnamefont {Omran}},
  \bibinfo {author} {\bibfnamefont {H.}~\bibnamefont {Levine}}, \bibinfo
  {author} {\bibfnamefont {H.}~\bibnamefont {Bernien}}, \bibinfo {author}
  {\bibfnamefont {H.}~\bibnamefont {Pichler}}, \bibinfo {author} {\bibfnamefont
  {S.}~\bibnamefont {Choi}}, \bibinfo {author} {\bibfnamefont {R.}~\bibnamefont
  {Samajdar}}, \bibinfo {author} {\bibfnamefont {S.}~\bibnamefont {Schwartz}},
  \bibinfo {author} {\bibfnamefont {P.}~\bibnamefont {Silvi}}, \bibinfo
  {author} {\bibfnamefont {S.}~\bibnamefont {Sachdev}}, \bibinfo {author}
  {\bibfnamefont {P.}~\bibnamefont {Zoller}}, \bibinfo {author} {\bibfnamefont
  {M.}~\bibnamefont {Endres}}, \bibinfo {author} {\bibfnamefont
  {M.}~\bibnamefont {Greiner}}, \bibinfo {author} {\bibfnamefont
  {V.}~\bibnamefont {Vuleti{\'c}}},\ and\ \bibinfo {author} {\bibfnamefont
  {M.~D.}\ \bibnamefont {Lukin}},\ }\bibfield  {title} {\bibinfo {title}
  {{Quantum Kibble--Zurek mechanism and critical dynamics on a programmable
  Rydberg simulator}},\ }\href {https://doi.org/10.1038/s41586-019-1070-1}
  {\bibfield  {journal} {\bibinfo  {journal} {Nature}\ }\textbf {\bibinfo
  {volume} {568}},\ \bibinfo {pages} {207} (\bibinfo {year}
  {2019})}\BibitemShut {NoStop}%
\bibitem [{\citenamefont {Bluvstein}\ \emph {et~al.}(2021)\citenamefont
  {Bluvstein}, \citenamefont {Omran}, \citenamefont {Levine}, \citenamefont
  {Keesling}, \citenamefont {Semeghini}, \citenamefont {Ebadi}, \citenamefont
  {Wang}, \citenamefont {Michailidis}, \citenamefont {Maskara}, \citenamefont
  {Ho}, \citenamefont {Choi}, \citenamefont {Serbyn}, \citenamefont {Greiner},
  \citenamefont {Vuleti\'c},\ and\ \citenamefont
  {Lukin}}]{bluvstein2021controlling}%
  \BibitemOpen
  \bibfield  {author} {\bibinfo {author} {\bibfnamefont {D.}~\bibnamefont
  {Bluvstein}}, \bibinfo {author} {\bibfnamefont {A.}~\bibnamefont {Omran}},
  \bibinfo {author} {\bibfnamefont {H.}~\bibnamefont {Levine}}, \bibinfo
  {author} {\bibfnamefont {A.}~\bibnamefont {Keesling}}, \bibinfo {author}
  {\bibfnamefont {G.}~\bibnamefont {Semeghini}}, \bibinfo {author}
  {\bibfnamefont {S.}~\bibnamefont {Ebadi}}, \bibinfo {author} {\bibfnamefont
  {T.~T.}\ \bibnamefont {Wang}}, \bibinfo {author} {\bibfnamefont {A.~A.}\
  \bibnamefont {Michailidis}}, \bibinfo {author} {\bibfnamefont
  {N.}~\bibnamefont {Maskara}}, \bibinfo {author} {\bibfnamefont {W.~W.}\
  \bibnamefont {Ho}}, \bibinfo {author} {\bibfnamefont {S.}~\bibnamefont
  {Choi}}, \bibinfo {author} {\bibfnamefont {M.}~\bibnamefont {Serbyn}},
  \bibinfo {author} {\bibfnamefont {M.}~\bibnamefont {Greiner}}, \bibinfo
  {author} {\bibfnamefont {V.}~\bibnamefont {Vuleti\'c}},\ and\ \bibinfo
  {author} {\bibfnamefont {M.~D.}\ \bibnamefont {Lukin}},\ }\bibfield  {title}
  {\bibinfo {title} {{Controlling quantum many-body dynamics in driven Rydberg
  atom arrays}},\ }\href {https://doi.org/10.1126/science.abg2530} {\bibfield
  {journal} {\bibinfo  {journal} {Science}\ }\textbf {\bibinfo {volume}
  {371}},\ \bibinfo {pages} {1355} (\bibinfo {year} {2021})}\BibitemShut
  {NoStop}%
\bibitem [{\citenamefont {Celi}\ \emph {et~al.}(2020)\citenamefont {Celi},
  \citenamefont {Vermersch}, \citenamefont {Viyuela}, \citenamefont {Pichler},
  \citenamefont {Lukin},\ and\ \citenamefont {Zoller}}]{Celi1}%
  \BibitemOpen
  \bibfield  {author} {\bibinfo {author} {\bibfnamefont {A.}~\bibnamefont
  {Celi}}, \bibinfo {author} {\bibfnamefont {B.}~\bibnamefont {Vermersch}},
  \bibinfo {author} {\bibfnamefont {O.}~\bibnamefont {Viyuela}}, \bibinfo
  {author} {\bibfnamefont {H.}~\bibnamefont {Pichler}}, \bibinfo {author}
  {\bibfnamefont {M.~D.}\ \bibnamefont {Lukin}},\ and\ \bibinfo {author}
  {\bibfnamefont {P.}~\bibnamefont {Zoller}},\ }\bibfield  {title} {\bibinfo
  {title} {{Emerging Two-Dimensional Gauge Theories in Rydberg Configurable
  Arrays}},\ }\href {https://doi.org/10.1103/PhysRevX.10.021057} {\bibfield
  {journal} {\bibinfo  {journal} {Phys. Rev. X}\ }\textbf {\bibinfo {volume}
  {10}},\ \bibinfo {pages} {021057} (\bibinfo {year} {2020})}\BibitemShut
  {NoStop}%
\bibitem [{\citenamefont {Surace}\ \emph {et~al.}(2020)\citenamefont {Surace},
  \citenamefont {Mazza}, \citenamefont {Giudici}, \citenamefont {Lerose},
  \citenamefont {Gambassi},\ and\ \citenamefont {Dalmonte}}]{Surace1}%
  \BibitemOpen
  \bibfield  {author} {\bibinfo {author} {\bibfnamefont {F.~M.}\ \bibnamefont
  {Surace}}, \bibinfo {author} {\bibfnamefont {P.~P.}\ \bibnamefont {Mazza}},
  \bibinfo {author} {\bibfnamefont {G.}~\bibnamefont {Giudici}}, \bibinfo
  {author} {\bibfnamefont {A.}~\bibnamefont {Lerose}}, \bibinfo {author}
  {\bibfnamefont {A.}~\bibnamefont {Gambassi}},\ and\ \bibinfo {author}
  {\bibfnamefont {M.}~\bibnamefont {Dalmonte}},\ }\bibfield  {title} {\bibinfo
  {title} {{Lattice Gauge Theories and String Dynamics in Rydberg Atom Quantum
  Simulators}},\ }\href {https://doi.org/10.1103/PhysRevX.10.021041} {\bibfield
   {journal} {\bibinfo  {journal} {Phys. Rev. X}\ }\textbf {\bibinfo {volume}
  {10}},\ \bibinfo {pages} {021041} (\bibinfo {year} {2020})}\BibitemShut
  {NoStop}%
\bibitem [{\citenamefont {Gonz\'alez-Cuadra}\ \emph {et~al.}(2022)\citenamefont
  {Gonz\'alez-Cuadra}, \citenamefont {Zache}, \citenamefont {Carrasco},
  \citenamefont {Kraus},\ and\ \citenamefont {Zoller}}]{qudit}%
  \BibitemOpen
  \bibfield  {author} {\bibinfo {author} {\bibfnamefont {D.}~\bibnamefont
  {Gonz\'alez-Cuadra}}, \bibinfo {author} {\bibfnamefont {T.~V.}\ \bibnamefont
  {Zache}}, \bibinfo {author} {\bibfnamefont {J.}~\bibnamefont {Carrasco}},
  \bibinfo {author} {\bibfnamefont {B.}~\bibnamefont {Kraus}},\ and\ \bibinfo
  {author} {\bibfnamefont {P.}~\bibnamefont {Zoller}},\ }\bibfield  {title}
  {\bibinfo {title} {{Hardware Efficient Quantum Simulation of Non-Abelian
  Gauge Theories with Qudits on Rydberg Platforms}},\ }\href
  {https://doi.org/10.1103/PhysRevLett.129.160501} {\bibfield  {journal}
  {\bibinfo  {journal} {Phys. Rev. Lett.}\ }\textbf {\bibinfo {volume} {129}},\
  \bibinfo {pages} {160501} (\bibinfo {year} {2022})}\BibitemShut {NoStop}%
\bibitem [{\citenamefont {Samajdar}\ \emph {et~al.}(2023)\citenamefont
  {Samajdar}, \citenamefont {Joshi}, \citenamefont {Teng},\ and\ \citenamefont
  {Sachdev}}]{IGT}%
  \BibitemOpen
  \bibfield  {author} {\bibinfo {author} {\bibfnamefont {R.}~\bibnamefont
  {Samajdar}}, \bibinfo {author} {\bibfnamefont {D.~G.}\ \bibnamefont {Joshi}},
  \bibinfo {author} {\bibfnamefont {Y.}~\bibnamefont {Teng}},\ and\ \bibinfo
  {author} {\bibfnamefont {S.}~\bibnamefont {Sachdev}},\ }\bibfield  {title}
  {\bibinfo {title} {{Emergent ${\mathbb{Z}}_{2}$ Gauge Theories and
  Topological Excitations in Rydberg Atom Arrays}},\ }\href
  {https://doi.org/10.1103/PhysRevLett.130.043601} {\bibfield  {journal}
  {\bibinfo  {journal} {Phys. Rev. Lett.}\ }\textbf {\bibinfo {volume} {130}},\
  \bibinfo {pages} {043601} (\bibinfo {year} {2023})}\BibitemShut {NoStop}%
\bibitem [{\citenamefont {Pichler}\ \emph {et~al.}(2018)\citenamefont
  {Pichler}, \citenamefont {Wang}, \citenamefont {Zhou}, \citenamefont {Choi},\
  and\ \citenamefont {Lukin}}]{pichler2018quantum}%
  \BibitemOpen
  \bibfield  {author} {\bibinfo {author} {\bibfnamefont {H.}~\bibnamefont
  {Pichler}}, \bibinfo {author} {\bibfnamefont {S.-T.}\ \bibnamefont {Wang}},
  \bibinfo {author} {\bibfnamefont {L.}~\bibnamefont {Zhou}}, \bibinfo {author}
  {\bibfnamefont {S.}~\bibnamefont {Choi}},\ and\ \bibinfo {author}
  {\bibfnamefont {M.~D.}\ \bibnamefont {Lukin}},\ }\bibfield  {title} {\bibinfo
  {title} {{Quantum optimization for maximum independent set using Rydberg atom
  arrays}},\ }\href {https://arxiv.org/abs/1808.10816} {\bibfield  {journal}
  {\bibinfo  {journal} {arXiv:1808.10816 [quant-ph]}\ } (\bibinfo {year}
  {2018})}\BibitemShut {NoStop}%
\bibitem [{\citenamefont {Ebadi}\ \emph {et~al.}(2022)\citenamefont {Ebadi},
  \citenamefont {Keesling}, \citenamefont {Cain}, \citenamefont {Wang},
  \citenamefont {Levine}, \citenamefont {Bluvstein}, \citenamefont {Semeghini},
  \citenamefont {Omran}, \citenamefont {Liu}, \citenamefont {Samajdar},
  \citenamefont {Luo}, \citenamefont {Nash}, \citenamefont {Gao}, \citenamefont
  {Barak}, \citenamefont {Farhi}, \citenamefont {Sachdev}, \citenamefont
  {Gemelke}, \citenamefont {Zhou}, \citenamefont {Choi}, \citenamefont
  {Pichler}, \citenamefont {Wang}, \citenamefont {Greiner}, \citenamefont
  {Vuletic},\ and\ \citenamefont {Lukin}}]{Ebadi2022}%
  \BibitemOpen
  \bibfield  {author} {\bibinfo {author} {\bibfnamefont {S.}~\bibnamefont
  {Ebadi}}, \bibinfo {author} {\bibfnamefont {A.}~\bibnamefont {Keesling}},
  \bibinfo {author} {\bibfnamefont {M.}~\bibnamefont {Cain}}, \bibinfo {author}
  {\bibfnamefont {T.~T.}\ \bibnamefont {Wang}}, \bibinfo {author}
  {\bibfnamefont {H.}~\bibnamefont {Levine}}, \bibinfo {author} {\bibfnamefont
  {D.}~\bibnamefont {Bluvstein}}, \bibinfo {author} {\bibfnamefont
  {G.}~\bibnamefont {Semeghini}}, \bibinfo {author} {\bibfnamefont
  {A.}~\bibnamefont {Omran}}, \bibinfo {author} {\bibfnamefont {J.-G.}\
  \bibnamefont {Liu}}, \bibinfo {author} {\bibfnamefont {R.}~\bibnamefont
  {Samajdar}}, \bibinfo {author} {\bibfnamefont {X.-Z.}\ \bibnamefont {Luo}},
  \bibinfo {author} {\bibfnamefont {B.}~\bibnamefont {Nash}}, \bibinfo {author}
  {\bibfnamefont {X.}~\bibnamefont {Gao}}, \bibinfo {author} {\bibfnamefont
  {B.}~\bibnamefont {Barak}}, \bibinfo {author} {\bibfnamefont
  {E.}~\bibnamefont {Farhi}}, \bibinfo {author} {\bibfnamefont
  {S.}~\bibnamefont {Sachdev}}, \bibinfo {author} {\bibfnamefont
  {N.}~\bibnamefont {Gemelke}}, \bibinfo {author} {\bibfnamefont
  {L.}~\bibnamefont {Zhou}}, \bibinfo {author} {\bibfnamefont {S.}~\bibnamefont
  {Choi}}, \bibinfo {author} {\bibfnamefont {H.}~\bibnamefont {Pichler}},
  \bibinfo {author} {\bibfnamefont {S.-T.}\ \bibnamefont {Wang}}, \bibinfo
  {author} {\bibfnamefont {M.}~\bibnamefont {Greiner}}, \bibinfo {author}
  {\bibfnamefont {V.}~\bibnamefont {Vuletic}},\ and\ \bibinfo {author}
  {\bibfnamefont {M.~D.}\ \bibnamefont {Lukin}},\ }\bibfield  {title} {\bibinfo
  {title} {Quantum optimization of maximum independent set using rydberg atom
  arrays},\ }\href {https://doi.org/10.1126/science.abo6587} {\bibfield
  {journal} {\bibinfo  {journal} {Science}\ }\textbf {\bibinfo {volume}
  {376}},\ \bibinfo {pages} {1209} (\bibinfo {year} {2022})}\BibitemShut
  {NoStop}%
\bibitem [{\citenamefont {Yan}\ \emph {et~al.}(2021{\natexlab{a}})\citenamefont
  {Yan}, \citenamefont {Zhou}, \citenamefont {Wang}, \citenamefont {Meng},\
  and\ \citenamefont {Zhang}}]{yan2021sweeping}%
  \BibitemOpen
  \bibfield  {author} {\bibinfo {author} {\bibfnamefont {Z.}~\bibnamefont
  {Yan}}, \bibinfo {author} {\bibfnamefont {Z.}~\bibnamefont {Zhou}}, \bibinfo
  {author} {\bibfnamefont {Y.-C.}\ \bibnamefont {Wang}}, \bibinfo {author}
  {\bibfnamefont {Z.~Y.}\ \bibnamefont {Meng}},\ and\ \bibinfo {author}
  {\bibfnamefont {X.-F.}\ \bibnamefont {Zhang}},\ }\bibfield  {title} {\bibinfo
  {title} {Preparing state within target topological sector of lattice gauge
  theory model on quantum simulator},\ }\href
  {https://arxiv.org/abs/2105.07134} {\bibfield  {journal} {\bibinfo  {journal}
  {arXiv:2105.07134 [quant-ph]}\ } (\bibinfo {year}
  {2021}{\natexlab{a}})}\BibitemShut {NoStop}%
\bibitem [{\citenamefont {Samajdar}\ \emph {et~al.}(2018)\citenamefont
  {Samajdar}, \citenamefont {Choi}, \citenamefont {Pichler}, \citenamefont
  {Lukin},\ and\ \citenamefont {Sachdev}}]{samajdar2018numerical}%
  \BibitemOpen
  \bibfield  {author} {\bibinfo {author} {\bibfnamefont {R.}~\bibnamefont
  {Samajdar}}, \bibinfo {author} {\bibfnamefont {S.}~\bibnamefont {Choi}},
  \bibinfo {author} {\bibfnamefont {H.}~\bibnamefont {Pichler}}, \bibinfo
  {author} {\bibfnamefont {M.~D.}\ \bibnamefont {Lukin}},\ and\ \bibinfo
  {author} {\bibfnamefont {S.}~\bibnamefont {Sachdev}},\ }\bibfield  {title}
  {\bibinfo {title} {Numerical study of the chiral $\mathbb{Z}_3$ quantum phase
  transition in one spatial dimension},\ }\href
  {https://doi.org/10.1103/PhysRevA.98.023614} {\bibfield  {journal} {\bibinfo
  {journal} {Phys. Rev. A}\ }\textbf {\bibinfo {volume} {98}},\ \bibinfo
  {pages} {023614} (\bibinfo {year} {2018})}\BibitemShut {NoStop}%
\bibitem [{\citenamefont {Whitsitt}\ \emph {et~al.}(2018)\citenamefont
  {Whitsitt}, \citenamefont {Samajdar},\ and\ \citenamefont
  {Sachdev}}]{whitsitt2018quantum}%
  \BibitemOpen
  \bibfield  {author} {\bibinfo {author} {\bibfnamefont {S.}~\bibnamefont
  {Whitsitt}}, \bibinfo {author} {\bibfnamefont {R.}~\bibnamefont {Samajdar}},\
  and\ \bibinfo {author} {\bibfnamefont {S.}~\bibnamefont {Sachdev}},\
  }\bibfield  {title} {\bibinfo {title} {Quantum field theory for the chiral
  clock transition in one spatial dimension},\ }\href
  {https://doi.org/10.1103/PhysRevB.98.205118} {\bibfield  {journal} {\bibinfo
  {journal} {Phys. Rev. B}\ }\textbf {\bibinfo {volume} {98}},\ \bibinfo
  {pages} {205118} (\bibinfo {year} {2018})}\BibitemShut {NoStop}%
\bibitem [{\citenamefont {Chepiga}\ and\ \citenamefont
  {Mila}(2019)}]{PhysRevLett.122.017205}%
  \BibitemOpen
  \bibfield  {author} {\bibinfo {author} {\bibfnamefont {N.}~\bibnamefont
  {Chepiga}}\ and\ \bibinfo {author} {\bibfnamefont {F.}~\bibnamefont {Mila}},\
  }\bibfield  {title} {\bibinfo {title} {{Floating Phase versus Chiral
  Transition in a 1D Hard-Boson Model}},\ }\href
  {https://doi.org/10.1103/PhysRevLett.122.017205} {\bibfield  {journal}
  {\bibinfo  {journal} {Phys. Rev. Lett.}\ }\textbf {\bibinfo {volume} {122}},\
  \bibinfo {pages} {017205} (\bibinfo {year} {2019})}\BibitemShut {NoStop}%
\bibitem [{\citenamefont {Chepiga}\ and\ \citenamefont
  {Mila}(2021)}]{chepiga2021kibble}%
  \BibitemOpen
  \bibfield  {author} {\bibinfo {author} {\bibfnamefont {N.}~\bibnamefont
  {Chepiga}}\ and\ \bibinfo {author} {\bibfnamefont {F.}~\bibnamefont {Mila}},\
  }\bibfield  {title} {\bibinfo {title} {{Kibble-Zurek exponent and chiral
  transition of the period-4 phase of Rydberg chains}},\ }\href
  {https://doi.org/10.1038/s41467-020-20641-y} {\bibfield  {journal} {\bibinfo
  {journal} {Nat. Commun.}\ }\textbf {\bibinfo {volume} {12}},\ \bibinfo
  {pages} {1} (\bibinfo {year} {2021})}\BibitemShut {NoStop}%
\bibitem [{\citenamefont {Samajdar}\ \emph {et~al.}(2020)\citenamefont
  {Samajdar}, \citenamefont {Ho}, \citenamefont {Pichler}, \citenamefont
  {Lukin},\ and\ \citenamefont {Sachdev}}]{Samajdar_2020}%
  \BibitemOpen
  \bibfield  {author} {\bibinfo {author} {\bibfnamefont {R.}~\bibnamefont
  {Samajdar}}, \bibinfo {author} {\bibfnamefont {W.~W.}\ \bibnamefont {Ho}},
  \bibinfo {author} {\bibfnamefont {H.}~\bibnamefont {Pichler}}, \bibinfo
  {author} {\bibfnamefont {M.~D.}\ \bibnamefont {Lukin}},\ and\ \bibinfo
  {author} {\bibfnamefont {S.}~\bibnamefont {Sachdev}},\ }\bibfield  {title}
  {\bibinfo {title} {{Complex Density Wave Orders and Quantum Phase Transitions
  in a Model of Square-Lattice Rydberg Atom Arrays}},\ }\href
  {https://doi.org/10.1103/physrevlett.124.103601} {\bibfield  {journal}
  {\bibinfo  {journal} {Phys. Rev. Lett.}\ }\textbf {\bibinfo {volume} {124}},\
  \bibinfo {pages} {103601} (\bibinfo {year} {2020})}\BibitemShut {NoStop}%
\bibitem [{\citenamefont {Felser}\ \emph {et~al.}(2021)\citenamefont {Felser},
  \citenamefont {Notarnicola},\ and\ \citenamefont
  {Montangero}}]{PhysRevLett.126.170603}%
  \BibitemOpen
  \bibfield  {author} {\bibinfo {author} {\bibfnamefont {T.}~\bibnamefont
  {Felser}}, \bibinfo {author} {\bibfnamefont {S.}~\bibnamefont
  {Notarnicola}},\ and\ \bibinfo {author} {\bibfnamefont {S.}~\bibnamefont
  {Montangero}},\ }\bibfield  {title} {\bibinfo {title} {{Efficient Tensor
  Network Ansatz for High-Dimensional Quantum Many-Body Problems}},\ }\href
  {https://doi.org/10.1103/PhysRevLett.126.170603} {\bibfield  {journal}
  {\bibinfo  {journal} {Phys. Rev. Lett.}\ }\textbf {\bibinfo {volume} {126}},\
  \bibinfo {pages} {170603} (\bibinfo {year} {2021})}\BibitemShut {NoStop}%
\bibitem [{\citenamefont {Kalinowski}\ \emph {et~al.}(2022)\citenamefont
  {Kalinowski}, \citenamefont {Samajdar}, \citenamefont {Melko}, \citenamefont
  {Lukin}, \citenamefont {Sachdev},\ and\ \citenamefont
  {Choi}}]{kalinowski2021bulk}%
  \BibitemOpen
  \bibfield  {author} {\bibinfo {author} {\bibfnamefont {M.}~\bibnamefont
  {Kalinowski}}, \bibinfo {author} {\bibfnamefont {R.}~\bibnamefont
  {Samajdar}}, \bibinfo {author} {\bibfnamefont {R.~G.}\ \bibnamefont {Melko}},
  \bibinfo {author} {\bibfnamefont {M.~D.}\ \bibnamefont {Lukin}}, \bibinfo
  {author} {\bibfnamefont {S.}~\bibnamefont {Sachdev}},\ and\ \bibinfo {author}
  {\bibfnamefont {S.}~\bibnamefont {Choi}},\ }\bibfield  {title} {\bibinfo
  {title} {{Bulk and boundary quantum phase transitions in a square Rydberg
  atom array}},\ }\href {https://doi.org/10.1103/PhysRevB.105.174417}
  {\bibfield  {journal} {\bibinfo  {journal} {Phys. Rev. B}\ }\textbf {\bibinfo
  {volume} {105}},\ \bibinfo {pages} {174417} (\bibinfo {year}
  {2022})}\BibitemShut {NoStop}%
\bibitem [{\citenamefont {Miles}\ \emph {et~al.}(2023)\citenamefont {Miles},
  \citenamefont {Samajdar}, \citenamefont {Ebadi}, \citenamefont {Wang},
  \citenamefont {Pichler}, \citenamefont {Sachdev}, \citenamefont {Lukin},
  \citenamefont {Greiner}, \citenamefont {Weinberger},\ and\ \citenamefont
  {Kim}}]{Kim.2021}%
  \BibitemOpen
  \bibfield  {author} {\bibinfo {author} {\bibfnamefont {C.}~\bibnamefont
  {Miles}}, \bibinfo {author} {\bibfnamefont {R.}~\bibnamefont {Samajdar}},
  \bibinfo {author} {\bibfnamefont {S.}~\bibnamefont {Ebadi}}, \bibinfo
  {author} {\bibfnamefont {T.~T.}\ \bibnamefont {Wang}}, \bibinfo {author}
  {\bibfnamefont {H.}~\bibnamefont {Pichler}}, \bibinfo {author} {\bibfnamefont
  {S.}~\bibnamefont {Sachdev}}, \bibinfo {author} {\bibfnamefont {M.~D.}\
  \bibnamefont {Lukin}}, \bibinfo {author} {\bibfnamefont {M.}~\bibnamefont
  {Greiner}}, \bibinfo {author} {\bibfnamefont {K.~Q.}\ \bibnamefont
  {Weinberger}},\ and\ \bibinfo {author} {\bibfnamefont {E.-A.}\ \bibnamefont
  {Kim}},\ }\bibfield  {title} {\bibinfo {title} {Machine learning discovery of
  new phases in programmable quantum simulator snapshots},\ }\href
  {https://doi.org/10.1103/PhysRevResearch.5.013026} {\bibfield  {journal}
  {\bibinfo  {journal} {Phys. Rev. Res.}\ }\textbf {\bibinfo {volume} {5}},\
  \bibinfo {pages} {013026} (\bibinfo {year} {2023})}\BibitemShut {NoStop}%
\bibitem [{\citenamefont {O'Rourke}\ and\ \citenamefont
  {Chan}(2022)}]{orourke2022entanglement}%
  \BibitemOpen
  \bibfield  {author} {\bibinfo {author} {\bibfnamefont {M.~J.}\ \bibnamefont
  {O'Rourke}}\ and\ \bibinfo {author} {\bibfnamefont {G.~K.-L.}\ \bibnamefont
  {Chan}},\ }\bibfield  {title} {\bibinfo {title} {{Entanglement in the quantum
  phases of an unfrustrated Rydberg atom array}},\ }\href
  {https://arxiv.org/abs/2201.03189} {\bibfield  {journal} {\bibinfo  {journal}
  {arXiv:2201.03189 [cond-mat.str-el]}\ } (\bibinfo {year} {2022})}\BibitemShut
  {NoStop}%
\bibitem [{\citenamefont {Li}\ \emph {et~al.}(2022)\citenamefont {Li},
  \citenamefont {Yang},\ and\ \citenamefont {Xu}}]{li2022quantum}%
  \BibitemOpen
  \bibfield  {author} {\bibinfo {author} {\bibfnamefont {C.-X.}\ \bibnamefont
  {Li}}, \bibinfo {author} {\bibfnamefont {S.}~\bibnamefont {Yang}},\ and\
  \bibinfo {author} {\bibfnamefont {J.-B.}\ \bibnamefont {Xu}},\ }\bibfield
  {title} {\bibinfo {title} {{Quantum phases of Rydberg atoms on a frustrated
  triangular-lattice array}},\ }\href {https://doi.org/10.1364/OL.450855}
  {\bibfield  {journal} {\bibinfo  {journal} {Opt. Lett.}\ }\textbf {\bibinfo
  {volume} {47}},\ \bibinfo {pages} {1093} (\bibinfo {year}
  {2022})}\BibitemShut {NoStop}%
\bibitem [{\citenamefont {Yang}\ and\ \citenamefont {Xu}(2022)}]{honeycomb}%
  \BibitemOpen
  \bibfield  {author} {\bibinfo {author} {\bibfnamefont {S.}~\bibnamefont
  {Yang}}\ and\ \bibinfo {author} {\bibfnamefont {J.-B.}\ \bibnamefont {Xu}},\
  }\bibfield  {title} {\bibinfo {title} {Density-wave-ordered phases of rydberg
  atoms on a honeycomb lattice},\ }\href
  {https://doi.org/10.1103/PhysRevE.106.034121} {\bibfield  {journal} {\bibinfo
   {journal} {Phys. Rev. E}\ }\textbf {\bibinfo {volume} {106}},\ \bibinfo
  {pages} {034121} (\bibinfo {year} {2022})}\BibitemShut {NoStop}%
\bibitem [{\citenamefont {Kornja{\v c}a}\ \emph {et~al.}(2022)\citenamefont
  {Kornja{\v c}a}, \citenamefont {Samajdar}, \citenamefont {Macr{\` i}},
  \citenamefont {Gemelke}, \citenamefont {Wang},\ and\ \citenamefont
  {Liu}}]{trimer}%
  \BibitemOpen
  \bibfield  {author} {\bibinfo {author} {\bibfnamefont {M.}~\bibnamefont
  {Kornja{\v c}a}}, \bibinfo {author} {\bibfnamefont {R.}~\bibnamefont
  {Samajdar}}, \bibinfo {author} {\bibfnamefont {T.}~\bibnamefont {Macr{\`
  i}}}, \bibinfo {author} {\bibfnamefont {N.}~\bibnamefont {Gemelke}}, \bibinfo
  {author} {\bibfnamefont {S.-T.}\ \bibnamefont {Wang}},\ and\ \bibinfo
  {author} {\bibfnamefont {F.}~\bibnamefont {Liu}},\ }\bibfield  {title}
  {\bibinfo {title} {{Trimer quantum spin liquid in a honeycomb array of
  Rydberg atoms}},\ }\href {https://arxiv.org/abs/2211.00653} {\bibfield
  {journal} {\bibinfo  {journal} {arXiv:2211.00653 [cond-mat.quant-gas]}\ }
  (\bibinfo {year} {2022})}\BibitemShut {NoStop}%
\bibitem [{\citenamefont {Samajdar}\ \emph {et~al.}(2021)\citenamefont
  {Samajdar}, \citenamefont {Ho}, \citenamefont {Pichler}, \citenamefont
  {Lukin},\ and\ \citenamefont {Sachdev}}]{Samajdar.2021}%
  \BibitemOpen
  \bibfield  {author} {\bibinfo {author} {\bibfnamefont {R.}~\bibnamefont
  {Samajdar}}, \bibinfo {author} {\bibfnamefont {W.~W.}\ \bibnamefont {Ho}},
  \bibinfo {author} {\bibfnamefont {H.}~\bibnamefont {Pichler}}, \bibinfo
  {author} {\bibfnamefont {M.~D.}\ \bibnamefont {Lukin}},\ and\ \bibinfo
  {author} {\bibfnamefont {S.}~\bibnamefont {Sachdev}},\ }\bibfield  {title}
  {\bibinfo {title} {{Quantum phases of Rydberg atoms on a kagome lattice}},\
  }\href {https://doi.org/10.1073/pnas.2015785118} {\bibfield  {journal}
  {\bibinfo  {journal} {Proc. Natl. Acad. Sci. U.S.A.}\ }\textbf {\bibinfo
  {volume} {118}},\ \bibinfo {pages} {e2015785118} (\bibinfo {year} {2021})},\
  \Eprint {https://arxiv.org/abs/2011.12295} {2011.12295} \BibitemShut
  {NoStop}%
\bibitem [{\citenamefont {Verresen}\ \emph {et~al.}(2021)\citenamefont
  {Verresen}, \citenamefont {Lukin},\ and\ \citenamefont
  {Vishwanath}}]{Verresen.2020}%
  \BibitemOpen
  \bibfield  {author} {\bibinfo {author} {\bibfnamefont {R.}~\bibnamefont
  {Verresen}}, \bibinfo {author} {\bibfnamefont {M.~D.}\ \bibnamefont
  {Lukin}},\ and\ \bibinfo {author} {\bibfnamefont {A.}~\bibnamefont
  {Vishwanath}},\ }\bibfield  {title} {\bibinfo {title} {{Prediction of Toric
  Code Topological Order from Rydberg Blockade}},\ }\href
  {https://doi.org/10.1103/PhysRevX.11.031005} {\bibfield  {journal} {\bibinfo
  {journal} {Phys. Rev. X}\ }\textbf {\bibinfo {volume} {11}},\ \bibinfo
  {pages} {031005} (\bibinfo {year} {2021})}\BibitemShut {NoStop}%
\bibitem [{\citenamefont {Giudici}\ \emph {et~al.}(2022)\citenamefont
  {Giudici}, \citenamefont {Lukin},\ and\ \citenamefont
  {Pichler}}]{hannes_dynamical}%
  \BibitemOpen
  \bibfield  {author} {\bibinfo {author} {\bibfnamefont {G.}~\bibnamefont
  {Giudici}}, \bibinfo {author} {\bibfnamefont {M.~D.}\ \bibnamefont {Lukin}},\
  and\ \bibinfo {author} {\bibfnamefont {H.}~\bibnamefont {Pichler}},\
  }\bibfield  {title} {\bibinfo {title} {{Dynamical Preparation of Quantum Spin
  Liquids in Rydberg Atom Arrays}},\ }\href
  {https://doi.org/10.1103/PhysRevLett.129.090401} {\bibfield  {journal}
  {\bibinfo  {journal} {Phys. Rev. Lett.}\ }\textbf {\bibinfo {volume} {129}},\
  \bibinfo {pages} {090401} (\bibinfo {year} {2022})}\BibitemShut {NoStop}%
\bibitem [{\citenamefont {Cheng}\ \emph {et~al.}(2021)\citenamefont {Cheng},
  \citenamefont {Li},\ and\ \citenamefont {Zhai}}]{zhai}%
  \BibitemOpen
  \bibfield  {author} {\bibinfo {author} {\bibfnamefont {Y.}~\bibnamefont
  {Cheng}}, \bibinfo {author} {\bibfnamefont {C.}~\bibnamefont {Li}},\ and\
  \bibinfo {author} {\bibfnamefont {H.}~\bibnamefont {Zhai}},\ }\bibfield
  {title} {\bibinfo {title} {{Variational Approach to Quantum Spin Liquid in a
  Rydberg Atom Simulator}},\ }\href {https://arxiv.org/abs/2112.13688}
  {\bibfield  {journal} {\bibinfo  {journal} {arXiv:2112.13688
  [cond-mat.quant-gas]}\ } (\bibinfo {year} {2021})}\BibitemShut {NoStop}%
\bibitem [{\citenamefont {Moessner}\ and\ \citenamefont
  {Raman}(2011)}]{moessner2011quantum}%
  \BibitemOpen
  \bibfield  {author} {\bibinfo {author} {\bibfnamefont {R.}~\bibnamefont
  {Moessner}}\ and\ \bibinfo {author} {\bibfnamefont {K.~S.}\ \bibnamefont
  {Raman}},\ }\bibfield  {title} {\bibinfo {title} {Quantum dimer models},\
  }in\ \href {https://doi.org/10.1007/978-3-642-10589-0_17} {\emph {\bibinfo
  {booktitle} {{Introduction to Frustrated Magnetism}}}}\ (\bibinfo
  {publisher} {Springer},\ \bibinfo {year} {2011})\ pp.\ \bibinfo {pages}
  {437--479}\BibitemShut {NoStop}%
\bibitem [{\citenamefont {Moessner}\ and\ \citenamefont
  {Sondhi}(2001)}]{moessner2001ising}%
  \BibitemOpen
  \bibfield  {author} {\bibinfo {author} {\bibfnamefont {R.}~\bibnamefont
  {Moessner}}\ and\ \bibinfo {author} {\bibfnamefont {S.~L.}\ \bibnamefont
  {Sondhi}},\ }\bibfield  {title} {\bibinfo {title} {Ising models of quantum
  frustration},\ }\href {https://doi.org/10.1103/PhysRevB.63.224401} {\bibfield
   {journal} {\bibinfo  {journal} {Phys. Rev. B}\ }\textbf {\bibinfo {volume}
  {63}},\ \bibinfo {pages} {224401} (\bibinfo {year} {2001})}\BibitemShut
  {NoStop}%
\bibitem [{\citenamefont {Roychowdhury}\ \emph {et~al.}(2015)\citenamefont
  {Roychowdhury}, \citenamefont {Bhattacharjee},\ and\ \citenamefont
  {Pollmann}}]{roychowdhury2015z}%
  \BibitemOpen
  \bibfield  {author} {\bibinfo {author} {\bibfnamefont {K.}~\bibnamefont
  {Roychowdhury}}, \bibinfo {author} {\bibfnamefont {S.}~\bibnamefont
  {Bhattacharjee}},\ and\ \bibinfo {author} {\bibfnamefont {F.}~\bibnamefont
  {Pollmann}},\ }\bibfield  {title} {\bibinfo {title} {{${\mathbb{Z}}_{2}$
  topological liquid of hard-core bosons on a kagome lattice at $1/3$
  filling}},\ }\href {https://doi.org/10.1103/PhysRevB.92.075141} {\bibfield
  {journal} {\bibinfo  {journal} {Phys. Rev. B}\ }\textbf {\bibinfo {volume}
  {92}},\ \bibinfo {pages} {075141} (\bibinfo {year} {2015})}\BibitemShut
  {NoStop}%
\bibitem [{\citenamefont {Yan}\ \emph {et~al.}(2022)\citenamefont {Yan},
  \citenamefont {Ran}, \citenamefont {Wang}, \citenamefont {Samajdar},
  \citenamefont {Rong}, \citenamefont {Sachdev}, \citenamefont {Qi},\ and\
  \citenamefont {Meng}}]{ZY2022loop}%
  \BibitemOpen
  \bibfield  {author} {\bibinfo {author} {\bibfnamefont {Z.}~\bibnamefont
  {Yan}}, \bibinfo {author} {\bibfnamefont {X.}~\bibnamefont {Ran}}, \bibinfo
  {author} {\bibfnamefont {Y.-C.}\ \bibnamefont {Wang}}, \bibinfo {author}
  {\bibfnamefont {R.}~\bibnamefont {Samajdar}}, \bibinfo {author}
  {\bibfnamefont {J.}~\bibnamefont {Rong}}, \bibinfo {author} {\bibfnamefont
  {S.}~\bibnamefont {Sachdev}}, \bibinfo {author} {\bibfnamefont
  {Y.}~\bibnamefont {Qi}},\ and\ \bibinfo {author} {\bibfnamefont {Z.~Y.}\
  \bibnamefont {Meng}},\ }\bibfield  {title} {\bibinfo {title} {Fully packed
  quantum loop model on the triangular lattice: Hidden vison plaquette phase
  and cubic phase transitions},\ }\href {https://arxiv.org/abs/2205.04472}
  {\bibfield  {journal} {\bibinfo  {journal} {arXiv:2205.04472
  [cond-mat.str-el]}\ } (\bibinfo {year} {2022})}\BibitemShut {NoStop}%
\bibitem [{\citenamefont {{Yan}}\ \emph {et~al.}(2022)\citenamefont {{Yan}},
  \citenamefont {{Samajdar}}, \citenamefont {{Wang}}, \citenamefont
  {{Sachdev}},\ and\ \citenamefont {{Meng}}}]{yan2022triangular}%
  \BibitemOpen
  \bibfield  {author} {\bibinfo {author} {\bibfnamefont {Z.}~\bibnamefont
  {{Yan}}}, \bibinfo {author} {\bibfnamefont {R.}~\bibnamefont {{Samajdar}}},
  \bibinfo {author} {\bibfnamefont {Y.-C.}\ \bibnamefont {{Wang}}}, \bibinfo
  {author} {\bibfnamefont {S.}~\bibnamefont {{Sachdev}}},\ and\ \bibinfo
  {author} {\bibfnamefont {Z.~Y.}\ \bibnamefont {{Meng}}},\ }\bibfield  {title}
  {\bibinfo {title} {{Triangular lattice quantum dimer model with variable
  dimer density}},\ }\href
  {https://doi.org/https://doi.org/10.1038/s41467-022-33431-5} {\bibfield
  {journal} {\bibinfo  {journal} {Nat. Commun.}\ }\textbf {\bibinfo {volume}
  {13}},\ \bibinfo {pages} {5799} (\bibinfo {year} {2022})}\BibitemShut
  {NoStop}%
\bibitem [{\citenamefont {Feldmeier}\ \emph {et~al.}(2019)\citenamefont
  {Feldmeier}, \citenamefont {Pollmann},\ and\ \citenamefont
  {Knap}}]{knap2019}%
  \BibitemOpen
  \bibfield  {author} {\bibinfo {author} {\bibfnamefont {J.}~\bibnamefont
  {Feldmeier}}, \bibinfo {author} {\bibfnamefont {F.}~\bibnamefont
  {Pollmann}},\ and\ \bibinfo {author} {\bibfnamefont {M.}~\bibnamefont
  {Knap}},\ }\bibfield  {title} {\bibinfo {title} {{Emergent Glassy Dynamics in
  a Quantum Dimer Model}},\ }\href
  {https://doi.org/10.1103/PhysRevLett.123.040601} {\bibfield  {journal}
  {\bibinfo  {journal} {Phys. Rev. Lett.}\ }\textbf {\bibinfo {volume} {123}},\
  \bibinfo {pages} {040601} (\bibinfo {year} {2019})}\BibitemShut {NoStop}%
\bibitem [{\citenamefont {Plat}\ \emph {et~al.}(2015)\citenamefont {Plat},
  \citenamefont {Alet}, \citenamefont {Capponi},\ and\ \citenamefont
  {Totsuka}}]{Plat2015z2}%
  \BibitemOpen
  \bibfield  {author} {\bibinfo {author} {\bibfnamefont {X.}~\bibnamefont
  {Plat}}, \bibinfo {author} {\bibfnamefont {F.}~\bibnamefont {Alet}}, \bibinfo
  {author} {\bibfnamefont {S.}~\bibnamefont {Capponi}},\ and\ \bibinfo {author}
  {\bibfnamefont {K.}~\bibnamefont {Totsuka}},\ }\bibfield  {title} {\bibinfo
  {title} {Magnetization plateaus of an easy-axis kagome antiferromagnet with
  extended interactions},\ }\href {https://doi.org/10.1103/PhysRevB.92.174402}
  {\bibfield  {journal} {\bibinfo  {journal} {Phys. Rev. B}\ }\textbf {\bibinfo
  {volume} {92}},\ \bibinfo {pages} {174402} (\bibinfo {year}
  {2015})}\BibitemShut {NoStop}%
\bibitem [{\citenamefont {{Moessner}}\ and\ \citenamefont
  {{Sondhi}}(2001)}]{RMSLS01}%
  \BibitemOpen
  \bibfield  {author} {\bibinfo {author} {\bibfnamefont {R.}~\bibnamefont
  {{Moessner}}}\ and\ \bibinfo {author} {\bibfnamefont {S.~L.}\ \bibnamefont
  {{Sondhi}}},\ }\bibfield  {title} {\bibinfo {title} {{Resonating Valence Bond
  Phase in the Triangular Lattice Quantum Dimer Model}},\ }\href
  {https://doi.org/10.1103/PhysRevLett.86.1881} {\bibfield  {journal} {\bibinfo
   {journal} {Phys. Rev. Lett.}\ }\textbf {\bibinfo {volume} {86}},\ \bibinfo
  {pages} {1881} (\bibinfo {year} {2001})}\BibitemShut {NoStop}%
\bibitem [{\citenamefont {Angelone}\ \emph {et~al.}(2016)\citenamefont
  {Angelone}, \citenamefont {Mezzacapo},\ and\ \citenamefont
  {Pupillo}}]{angelone2016superglass}%
  \BibitemOpen
  \bibfield  {author} {\bibinfo {author} {\bibfnamefont {A.}~\bibnamefont
  {Angelone}}, \bibinfo {author} {\bibfnamefont {F.}~\bibnamefont
  {Mezzacapo}},\ and\ \bibinfo {author} {\bibfnamefont {G.}~\bibnamefont
  {Pupillo}},\ }\bibfield  {title} {\bibinfo {title} {Superglass phase of
  interaction-blockaded gases on a triangular lattice},\ }\href
  {https://doi.org/10.1103/PhysRevLett.116.135303} {\bibfield  {journal}
  {\bibinfo  {journal} {Phys. Rev. Lett.}\ }\textbf {\bibinfo {volume} {116}},\
  \bibinfo {pages} {135303} (\bibinfo {year} {2016})}\BibitemShut {NoStop}%
\bibitem [{\citenamefont {Lesanovsky}\ and\ \citenamefont
  {Garrahan}(2013)}]{lesanovsky2013kinetic}%
  \BibitemOpen
  \bibfield  {author} {\bibinfo {author} {\bibfnamefont {I.}~\bibnamefont
  {Lesanovsky}}\ and\ \bibinfo {author} {\bibfnamefont {J.~P.}\ \bibnamefont
  {Garrahan}},\ }\bibfield  {title} {\bibinfo {title} {{Kinetic constraints,
  hierarchical relaxation, and onset of glassiness in strongly interacting and
  dissipative Rydberg gases}},\ }\href
  {https://doi.org/10.1103/PhysRevLett.111.215305} {\bibfield  {journal}
  {\bibinfo  {journal} {Phys. Rev. Lett.}\ }\textbf {\bibinfo {volume} {111}},\
  \bibinfo {pages} {215305} (\bibinfo {year} {2013})}\BibitemShut {NoStop}%
\bibitem [{\citenamefont {P\'erez-Espigares}\ \emph {et~al.}(2018)\citenamefont
  {P\'erez-Espigares}, \citenamefont {Lesanovsky}, \citenamefont {Garrahan},\
  and\ \citenamefont {Guti\'errez}}]{carlos2018}%
  \BibitemOpen
  \bibfield  {author} {\bibinfo {author} {\bibfnamefont {C.}~\bibnamefont
  {P\'erez-Espigares}}, \bibinfo {author} {\bibfnamefont {I.}~\bibnamefont
  {Lesanovsky}}, \bibinfo {author} {\bibfnamefont {J.~P.}\ \bibnamefont
  {Garrahan}},\ and\ \bibinfo {author} {\bibfnamefont {R.}~\bibnamefont
  {Guti\'errez}},\ }\bibfield  {title} {\bibinfo {title} {{Glassy dynamics due
  to a trajectory phase transition in dissipative Rydberg gases}},\ }\href
  {https://doi.org/10.1103/PhysRevA.98.021804} {\bibfield  {journal} {\bibinfo
  {journal} {Phys. Rev. A}\ }\textbf {\bibinfo {volume} {98}},\ \bibinfo
  {pages} {021804} (\bibinfo {year} {2018})}\BibitemShut {NoStop}%
\bibitem [{\citenamefont {Sandvik}(2003)}]{Sandvik2003}%
  \BibitemOpen
  \bibfield  {author} {\bibinfo {author} {\bibfnamefont {A.~W.}\ \bibnamefont
  {Sandvik}},\ }\bibfield  {title} {\bibinfo {title} {Stochastic series
  expansion method for quantum {I}sing models with arbitrary interactions},\
  }\href {https://doi.org/10.1103/PhysRevE.68.056701} {\bibfield  {journal}
  {\bibinfo  {journal} {Phys. Rev. E}\ }\textbf {\bibinfo {volume} {68}},\
  \bibinfo {pages} {056701} (\bibinfo {year} {2003})}\BibitemShut {NoStop}%
\bibitem [{\citenamefont {Yan}\ \emph {et~al.}(2019)\citenamefont {Yan},
  \citenamefont {Wu}, \citenamefont {Liu}, \citenamefont {Sylju\aa{}sen},
  \citenamefont {Lou},\ and\ \citenamefont {Chen}}]{ZY2019sweeping}%
  \BibitemOpen
  \bibfield  {author} {\bibinfo {author} {\bibfnamefont {Z.}~\bibnamefont
  {Yan}}, \bibinfo {author} {\bibfnamefont {Y.}~\bibnamefont {Wu}}, \bibinfo
  {author} {\bibfnamefont {C.}~\bibnamefont {Liu}}, \bibinfo {author}
  {\bibfnamefont {O.~F.}\ \bibnamefont {Sylju\aa{}sen}}, \bibinfo {author}
  {\bibfnamefont {J.}~\bibnamefont {Lou}},\ and\ \bibinfo {author}
  {\bibfnamefont {Y.}~\bibnamefont {Chen}},\ }\bibfield  {title} {\bibinfo
  {title} {Sweeping cluster algorithm for quantum spin systems with strong
  geometric restrictions},\ }\href {https://doi.org/10.1103/PhysRevB.99.165135}
  {\bibfield  {journal} {\bibinfo  {journal} {Phys. Rev. B}\ }\textbf {\bibinfo
  {volume} {99}},\ \bibinfo {pages} {165135} (\bibinfo {year}
  {2019})}\BibitemShut {NoStop}%
\bibitem [{\citenamefont {Yan}(2022)}]{ZY2020improved}%
  \BibitemOpen
  \bibfield  {author} {\bibinfo {author} {\bibfnamefont {Z.}~\bibnamefont
  {Yan}},\ }\bibfield  {title} {\bibinfo {title} {Global scheme of sweeping
  cluster algorithm to sample among topological sectors},\ }\href
  {https://doi.org/10.1103/PhysRevB.105.184432} {\bibfield  {journal} {\bibinfo
   {journal} {Phys. Rev. B}\ }\textbf {\bibinfo {volume} {105}},\ \bibinfo
  {pages} {184432} (\bibinfo {year} {2022})}\BibitemShut {NoStop}%
\bibitem [{\citenamefont {Yan}\ \emph {et~al.}(2021{\natexlab{b}})\citenamefont
  {Yan}, \citenamefont {Wang}, \citenamefont {Ma}, \citenamefont {Qi},\ and\
  \citenamefont {Meng}}]{ZY2020}%
  \BibitemOpen
  \bibfield  {author} {\bibinfo {author} {\bibfnamefont {Z.}~\bibnamefont
  {Yan}}, \bibinfo {author} {\bibfnamefont {Y.-C.}\ \bibnamefont {Wang}},
  \bibinfo {author} {\bibfnamefont {N.}~\bibnamefont {Ma}}, \bibinfo {author}
  {\bibfnamefont {Y.}~\bibnamefont {Qi}},\ and\ \bibinfo {author}
  {\bibfnamefont {Z.~Y.}\ \bibnamefont {Meng}},\ }\bibfield  {title} {\bibinfo
  {title} {Topological phase transition and single/multi anyon dynamics of
  ${Z}_2$ spin liquid},\ }\href {https://doi.org/10.1038/s41535-021-00338-1}
  {\bibfield  {journal} {\bibinfo  {journal} {npj Quantum Mater.}\ }\textbf
  {\bibinfo {volume} {6}},\ \bibinfo {pages} {39} (\bibinfo {year}
  {2021}{\natexlab{b}})}\BibitemShut {NoStop}%
\bibitem [{\citenamefont {Yan}\ \emph {et~al.}(2021{\natexlab{c}})\citenamefont
  {Yan}, \citenamefont {Zhou}, \citenamefont {Sylju\aa{}sen}, \citenamefont
  {Zhang}, \citenamefont {Yuan}, \citenamefont {Lou},\ and\ \citenamefont
  {Chen}}]{ZY2021mixed}%
  \BibitemOpen
  \bibfield  {author} {\bibinfo {author} {\bibfnamefont {Z.}~\bibnamefont
  {Yan}}, \bibinfo {author} {\bibfnamefont {Z.}~\bibnamefont {Zhou}}, \bibinfo
  {author} {\bibfnamefont {O.~F.}\ \bibnamefont {Sylju\aa{}sen}}, \bibinfo
  {author} {\bibfnamefont {J.}~\bibnamefont {Zhang}}, \bibinfo {author}
  {\bibfnamefont {T.}~\bibnamefont {Yuan}}, \bibinfo {author} {\bibfnamefont
  {J.}~\bibnamefont {Lou}},\ and\ \bibinfo {author} {\bibfnamefont
  {Y.}~\bibnamefont {Chen}},\ }\bibfield  {title} {\bibinfo {title} {Widely
  existing mixed phase structure of the quantum dimer model on a square
  lattice},\ }\href {https://doi.org/10.1103/PhysRevB.103.094421} {\bibfield
  {journal} {\bibinfo  {journal} {Phys. Rev. B}\ }\textbf {\bibinfo {volume}
  {103}},\ \bibinfo {pages} {094421} (\bibinfo {year}
  {2021}{\natexlab{c}})}\BibitemShut {NoStop}%
\bibitem [{\citenamefont {Da~Liao}\ \emph {et~al.}(2021)\citenamefont
  {Da~Liao}, \citenamefont {Li}, \citenamefont {Yan}, \citenamefont {Wei},
  \citenamefont {Li}, \citenamefont {Qi},\ and\ \citenamefont
  {Meng}}]{da2021phase}%
  \BibitemOpen
  \bibfield  {author} {\bibinfo {author} {\bibfnamefont {Y.}~\bibnamefont
  {Da~Liao}}, \bibinfo {author} {\bibfnamefont {H.}~\bibnamefont {Li}},
  \bibinfo {author} {\bibfnamefont {Z.}~\bibnamefont {Yan}}, \bibinfo {author}
  {\bibfnamefont {H.-T.}\ \bibnamefont {Wei}}, \bibinfo {author} {\bibfnamefont
  {W.}~\bibnamefont {Li}}, \bibinfo {author} {\bibfnamefont {Y.}~\bibnamefont
  {Qi}},\ and\ \bibinfo {author} {\bibfnamefont {Z.~Y.}\ \bibnamefont {Meng}},\
  }\bibfield  {title} {\bibinfo {title} {{Phase diagram of the quantum Ising
  model on a triangular lattice under external field}},\ }\href
  {https://doi.org/10.1103/PhysRevB.103.104416} {\bibfield  {journal} {\bibinfo
   {journal} {Phys. Rev. B}\ }\textbf {\bibinfo {volume} {103}},\ \bibinfo
  {pages} {104416} (\bibinfo {year} {2021})}\BibitemShut {NoStop}%
\bibitem [{\citenamefont {Merali}\ \emph {et~al.}(2021)\citenamefont {Merali},
  \citenamefont {De~Vlugt},\ and\ \citenamefont
  {Melko}}]{merali2021stochastic}%
  \BibitemOpen
  \bibfield  {author} {\bibinfo {author} {\bibfnamefont {E.}~\bibnamefont
  {Merali}}, \bibinfo {author} {\bibfnamefont {I.~J.}\ \bibnamefont
  {De~Vlugt}},\ and\ \bibinfo {author} {\bibfnamefont {R.~G.}\ \bibnamefont
  {Melko}},\ }\bibfield  {title} {\bibinfo {title} {{Stochastic Series
  Expansion Quantum Monte Carlo for Rydberg Arrays}},\ }\href
  {https://arxiv.org/abs/2107.00766} {\bibfield  {journal} {\bibinfo  {journal}
  {arXiv:2107.00766 [cond-mat.str-el]}\ } (\bibinfo {year} {2021})}\BibitemShut
  {NoStop}%
\bibitem [{\citenamefont {Hukushima}\ and\ \citenamefont
  {Nemoto}(1996)}]{hukushima1996exchange}%
  \BibitemOpen
  \bibfield  {author} {\bibinfo {author} {\bibfnamefont {K.}~\bibnamefont
  {Hukushima}}\ and\ \bibinfo {author} {\bibfnamefont {K.}~\bibnamefont
  {Nemoto}},\ }\bibfield  {title} {\bibinfo {title} {{Exchange Monte Carlo
  method and application to spin glass simulations}},\ }\href
  {https://doi.org/10.1143/JPSJ.65.1604} {\bibfield  {journal} {\bibinfo
  {journal} {J. Phys. Soc. Japan}\ }\textbf {\bibinfo {volume} {65}},\ \bibinfo
  {pages} {1604} (\bibinfo {year} {1996})}\BibitemShut {NoStop}%
\bibitem [{\citenamefont {Brooke}\ \emph {et~al.}(1999)\citenamefont {Brooke},
  \citenamefont {Bitko}, \citenamefont {Rosenbaum},\ and\ \citenamefont
  {Aeppli}}]{Brooke1999}%
  \BibitemOpen
  \bibfield  {author} {\bibinfo {author} {\bibfnamefont {J.}~\bibnamefont
  {Brooke}}, \bibinfo {author} {\bibfnamefont {D.}~\bibnamefont {Bitko}},
  \bibinfo {author} {\bibfnamefont {F.~T.}\ \bibnamefont {Rosenbaum}},\ and\
  \bibinfo {author} {\bibfnamefont {G.}~\bibnamefont {Aeppli}},\ }\bibfield
  {title} {\bibinfo {title} {Quantum annealing of a disordered magnet},\ }\href
  {https://doi.org/10.1126/science.284.5415.779} {\bibfield  {journal}
  {\bibinfo  {journal} {Science}\ }\textbf {\bibinfo {volume} {284}},\ \bibinfo
  {pages} {779} (\bibinfo {year} {1999})}\BibitemShut {NoStop}%
\bibitem [{\citenamefont {Brooke}\ \emph {et~al.}(2001)\citenamefont {Brooke},
  \citenamefont {Rosenbaum},\ and\ \citenamefont {Aeppli}}]{Brooke2001}%
  \BibitemOpen
  \bibfield  {author} {\bibinfo {author} {\bibfnamefont {J.}~\bibnamefont
  {Brooke}}, \bibinfo {author} {\bibfnamefont {T.}~\bibnamefont {Rosenbaum}},\
  and\ \bibinfo {author} {\bibfnamefont {G.}~\bibnamefont {Aeppli}},\
  }\bibfield  {title} {\bibinfo {title} {Tunable quantum tunnelling of magnetic
  domain walls},\ }\href {https://doi.org/10.1038/35098037} {\bibfield
  {journal} {\bibinfo  {journal} {Nature}\ }\textbf {\bibinfo {volume} {413}},\
  \bibinfo {pages} {610} (\bibinfo {year} {2001})}\BibitemShut {NoStop}%
\bibitem [{\citenamefont {Farhi}\ \emph {et~al.}(2000)\citenamefont {Farhi},
  \citenamefont {Goldstone}, \citenamefont {Gutmann},\ and\ \citenamefont
  {Sipser}}]{Sipser00}%
  \BibitemOpen
  \bibfield  {author} {\bibinfo {author} {\bibfnamefont {E.}~\bibnamefont
  {Farhi}}, \bibinfo {author} {\bibfnamefont {J.}~\bibnamefont {Goldstone}},
  \bibinfo {author} {\bibfnamefont {S.}~\bibnamefont {Gutmann}},\ and\ \bibinfo
  {author} {\bibfnamefont {M.}~\bibnamefont {Sipser}},\ }\href@noop {}
  {\bibinfo {title} {Quantum computation by adiabatic evolution}} (\bibinfo
  {year} {2000}),\ \Eprint {https://arxiv.org/abs/quant-ph/0001106}
  {arXiv:quant-ph/0001106} \BibitemShut {NoStop}%
\bibitem [{\citenamefont {Farhi}\ \emph {et~al.}(2001)\citenamefont {Farhi},
  \citenamefont {Goldstone}, \citenamefont {Gutmann}, \citenamefont {Lapan},
  \citenamefont {Lundgren},\ and\ \citenamefont {Preda}}]{Preda2001}%
  \BibitemOpen
  \bibfield  {author} {\bibinfo {author} {\bibfnamefont {E.}~\bibnamefont
  {Farhi}}, \bibinfo {author} {\bibfnamefont {J.}~\bibnamefont {Goldstone}},
  \bibinfo {author} {\bibfnamefont {S.}~\bibnamefont {Gutmann}}, \bibinfo
  {author} {\bibfnamefont {J.}~\bibnamefont {Lapan}}, \bibinfo {author}
  {\bibfnamefont {A.}~\bibnamefont {Lundgren}},\ and\ \bibinfo {author}
  {\bibfnamefont {D.}~\bibnamefont {Preda}},\ }\bibfield  {title} {\bibinfo
  {title} {A quantum adiabatic evolution algorithm applied to random instances
  of an {NP}-complete problem},\ }\href
  {https://doi.org/10.1126/science.1057726} {\bibfield  {journal} {\bibinfo
  {journal} {Science}\ }\textbf {\bibinfo {volume} {292}},\ \bibinfo {pages}
  {472} (\bibinfo {year} {2001})}\BibitemShut {NoStop}%
\bibitem [{\citenamefont {Melko}(2007)}]{melko2007simulations}%
  \BibitemOpen
  \bibfield  {author} {\bibinfo {author} {\bibfnamefont {R.~G.}\ \bibnamefont
  {Melko}},\ }\bibfield  {title} {\bibinfo {title} {{Simulations of quantum XXZ
  models on two-dimensional frustrated lattices}},\ }\href
  {https://doi.org/10.1088/0953-8984/19/14/145203} {\bibfield  {journal}
  {\bibinfo  {journal} {J. Phys.: Condens. Matter}\ }\textbf {\bibinfo {volume}
  {19}},\ \bibinfo {pages} {145203} (\bibinfo {year} {2007})}\BibitemShut
  {NoStop}%
\bibitem [{\citenamefont {Das}\ and\ \citenamefont
  {Chakrabarti}(2008)}]{Das2008}%
  \BibitemOpen
  \bibfield  {author} {\bibinfo {author} {\bibfnamefont {A.}~\bibnamefont
  {Das}}\ and\ \bibinfo {author} {\bibfnamefont {B.~K.}\ \bibnamefont
  {Chakrabarti}},\ }\bibfield  {title} {\bibinfo {title} {Colloquium: Quantum
  annealing and analog quantum computation},\ }\href
  {https://doi.org/10.1103/RevModPhys.80.1061} {\bibfield  {journal} {\bibinfo
  {journal} {Rev. Mod. Phys.}\ }\textbf {\bibinfo {volume} {80}},\ \bibinfo
  {pages} {1061} (\bibinfo {year} {2008})}\BibitemShut {NoStop}%
\bibitem [{\citenamefont {Aranson}\ \emph {et~al.}(2001)\citenamefont
  {Aranson}, \citenamefont {Kopnin},\ and\ \citenamefont
  {Vinokur}}]{PhysRevB.63.184501}%
  \BibitemOpen
  \bibfield  {author} {\bibinfo {author} {\bibfnamefont {I.~S.}\ \bibnamefont
  {Aranson}}, \bibinfo {author} {\bibfnamefont {N.~B.}\ \bibnamefont
  {Kopnin}},\ and\ \bibinfo {author} {\bibfnamefont {V.~M.}\ \bibnamefont
  {Vinokur}},\ }\bibfield  {title} {\bibinfo {title} {Dynamics of vortex
  nucleation by rapid thermal quench},\ }\href
  {https://doi.org/10.1103/PhysRevB.63.184501} {\bibfield  {journal} {\bibinfo
  {journal} {Phys. Rev. B}\ }\textbf {\bibinfo {volume} {63}},\ \bibinfo
  {pages} {184501} (\bibinfo {year} {2001})}\BibitemShut {NoStop}%
\bibitem [{\citenamefont {Mitra}(2018)}]{mitra2018quantum}%
  \BibitemOpen
  \bibfield  {author} {\bibinfo {author} {\bibfnamefont {A.}~\bibnamefont
  {Mitra}},\ }\bibfield  {title} {\bibinfo {title} {Quantum quench dynamics},\
  }\href {https://doi.org/10.1146/annurev-conmatphys-031016-025451} {\bibfield
  {journal} {\bibinfo  {journal} {Annu. Rev. Condens. Matter Phys.}\ }\textbf
  {\bibinfo {volume} {9}},\ \bibinfo {pages} {245} (\bibinfo {year}
  {2018})}\BibitemShut {NoStop}%
\bibitem [{\citenamefont {Edwards}\ and\ \citenamefont
  {Anderson}(1975)}]{Edwards1975}%
  \BibitemOpen
  \bibfield  {author} {\bibinfo {author} {\bibfnamefont {S.~F.}\ \bibnamefont
  {Edwards}}\ and\ \bibinfo {author} {\bibfnamefont {P.~W.}\ \bibnamefont
  {Anderson}},\ }\bibfield  {title} {\bibinfo {title} {Theory of spin
  glasses},\ }\href {https://doi.org/10.1088/0305-4608/5/5/017} {\bibfield
  {journal} {\bibinfo  {journal} {J. Phys. F}\ }\textbf {\bibinfo {volume}
  {5}},\ \bibinfo {pages} {965} (\bibinfo {year} {1975})}\BibitemShut {NoStop}%
\bibitem [{\citenamefont {Richards}(1984)}]{richards1984spin}%
  \BibitemOpen
  \bibfield  {author} {\bibinfo {author} {\bibfnamefont {P.~M.}\ \bibnamefont
  {Richards}},\ }\bibfield  {title} {\bibinfo {title} {{Spin-glass order
  parameter of the random-field Ising model}},\ }\href
  {https://doi.org/10.1103/PhysRevB.30.2955} {\bibfield  {journal} {\bibinfo
  {journal} {Phys. Rev. B}\ }\textbf {\bibinfo {volume} {30}},\ \bibinfo
  {pages} {2955} (\bibinfo {year} {1984})}\BibitemShut {NoStop}%
\bibitem [{\citenamefont {Georges}\ \emph {et~al.}(2001)\citenamefont
  {Georges}, \citenamefont {Parcollet},\ and\ \citenamefont
  {Sachdev}}]{georges2001quantum}%
  \BibitemOpen
  \bibfield  {author} {\bibinfo {author} {\bibfnamefont {A.}~\bibnamefont
  {Georges}}, \bibinfo {author} {\bibfnamefont {O.}~\bibnamefont {Parcollet}},\
  and\ \bibinfo {author} {\bibfnamefont {S.}~\bibnamefont {Sachdev}},\
  }\bibfield  {title} {\bibinfo {title} {{Quantum fluctuations of a nearly
  critical Heisenberg spin glass}},\ }\href
  {https://doi.org/10.1103/PhysRevB.63.134406} {\bibfield  {journal} {\bibinfo
  {journal} {Phys. Rev. B}\ }\textbf {\bibinfo {volume} {63}},\ \bibinfo
  {pages} {134406} (\bibinfo {year} {2001})}\BibitemShut {NoStop}%
\bibitem [{\citenamefont {Binder}\ and\ \citenamefont
  {Young}(1986)}]{binder1986spin}%
  \BibitemOpen
  \bibfield  {author} {\bibinfo {author} {\bibfnamefont {K.}~\bibnamefont
  {Binder}}\ and\ \bibinfo {author} {\bibfnamefont {A.~P.}\ \bibnamefont
  {Young}},\ }\bibfield  {title} {\bibinfo {title} {{Spin glasses: Experimental
  facts, theoretical concepts, and open questions}},\ }\href
  {https://doi.org/10.1103/RevModPhys.58.801} {\bibfield  {journal} {\bibinfo
  {journal} {Rev. Mod. Phys.}\ }\textbf {\bibinfo {volume} {58}},\ \bibinfo
  {pages} {801} (\bibinfo {year} {1986})}\BibitemShut {NoStop}%
\bibitem [{sup()}]{suppl}%
  \BibitemOpen
  \href@noop {} {\bibinfo  {journal} {The implementation of the different
  simulating schemes (annealing, quenching and parallel tempering), a brief
  description of the Edwards-Anderson order parameter, the phase transition
  between the nematic and paramagnetic phases, and the snapshots in the glassy
  region are presented in the Supplemental Material}\ }\BibitemShut {NoStop}%
\bibitem [{\citenamefont {Sachdev}(2011)}]{sachdev2011quantum}%
  \BibitemOpen
\bibfield  {journal} {  }\bibfield  {author} {\bibinfo {author} {\bibfnamefont
  {S.}~\bibnamefont {Sachdev}},\ }\href
  {https://doi.org/10.1017/CBO9780511973765} {\emph {\bibinfo {title} {Quantum
  {P}hase {T}ransitions}}}\ (\bibinfo  {publisher} {Cambridge University
  Press},\ \bibinfo {address} {New York},\ \bibinfo {year} {2011})\BibitemShut
  {NoStop}%
\bibitem [{\citenamefont {Janke}\ and\ \citenamefont
  {Villanova}(1997)}]{janke1997three}%
  \BibitemOpen
  \bibfield  {author} {\bibinfo {author} {\bibfnamefont {W.}~\bibnamefont
  {Janke}}\ and\ \bibinfo {author} {\bibfnamefont {R.}~\bibnamefont
  {Villanova}},\ }\bibfield  {title} {\bibinfo {title} {{Three-dimensional
  3-state Potts model revisited with new techniques}},\ }\href
  {https://doi.org/10.1016/S0550-3213(96)00710-9} {\bibfield  {journal}
  {\bibinfo  {journal} {Nucl. Phys. B}\ }\textbf {\bibinfo {volume} {489}},\
  \bibinfo {pages} {679} (\bibinfo {year} {1997})}\BibitemShut {NoStop}%
\bibitem [{\citenamefont {King}\ \emph {et~al.}(2022)\citenamefont {King},
  \citenamefont {Raymond}, \citenamefont {Lanting}, \citenamefont {Harris},
  \citenamefont {Zucca}, \citenamefont {Altomare}, \citenamefont {Berkley},
  \citenamefont {Boothby}, \citenamefont {Ejtemaee}, \citenamefont {Enderud},
  \citenamefont {Hoskinson}, \citenamefont {Huang}, \citenamefont {Ladizinsky},
  \citenamefont {MacDonald}, \citenamefont {Marsden}, \citenamefont {Molavi},
  \citenamefont {Oh}, \citenamefont {Poulin-Lamarre}, \citenamefont {Reis},
  \citenamefont {Rich}, \citenamefont {Sato}, \citenamefont {Tsai},
  \citenamefont {Volkmann}, \citenamefont {Whittaker}, \citenamefont {Yao},
  \citenamefont {Sandvik},\ and\ \citenamefont {Amin}}]{king2022quantum}%
  \BibitemOpen
  \bibfield  {author} {\bibinfo {author} {\bibfnamefont {A.~D.}\ \bibnamefont
  {King}}, \bibinfo {author} {\bibfnamefont {J.}~\bibnamefont {Raymond}},
  \bibinfo {author} {\bibfnamefont {T.}~\bibnamefont {Lanting}}, \bibinfo
  {author} {\bibfnamefont {R.}~\bibnamefont {Harris}}, \bibinfo {author}
  {\bibfnamefont {A.}~\bibnamefont {Zucca}}, \bibinfo {author} {\bibfnamefont
  {F.}~\bibnamefont {Altomare}}, \bibinfo {author} {\bibfnamefont {A.~J.}\
  \bibnamefont {Berkley}}, \bibinfo {author} {\bibfnamefont {K.}~\bibnamefont
  {Boothby}}, \bibinfo {author} {\bibfnamefont {S.}~\bibnamefont {Ejtemaee}},
  \bibinfo {author} {\bibfnamefont {C.}~\bibnamefont {Enderud}}, \bibinfo
  {author} {\bibfnamefont {E.}~\bibnamefont {Hoskinson}}, \bibinfo {author}
  {\bibfnamefont {S.}~\bibnamefont {Huang}}, \bibinfo {author} {\bibfnamefont
  {E.}~\bibnamefont {Ladizinsky}}, \bibinfo {author} {\bibfnamefont {A.~J.~R.}\
  \bibnamefont {MacDonald}}, \bibinfo {author} {\bibfnamefont {G.}~\bibnamefont
  {Marsden}}, \bibinfo {author} {\bibfnamefont {R.}~\bibnamefont {Molavi}},
  \bibinfo {author} {\bibfnamefont {T.}~\bibnamefont {Oh}}, \bibinfo {author}
  {\bibfnamefont {G.}~\bibnamefont {Poulin-Lamarre}}, \bibinfo {author}
  {\bibfnamefont {M.}~\bibnamefont {Reis}}, \bibinfo {author} {\bibfnamefont
  {C.}~\bibnamefont {Rich}}, \bibinfo {author} {\bibfnamefont {Y.}~\bibnamefont
  {Sato}}, \bibinfo {author} {\bibfnamefont {N.}~\bibnamefont {Tsai}}, \bibinfo
  {author} {\bibfnamefont {M.}~\bibnamefont {Volkmann}}, \bibinfo {author}
  {\bibfnamefont {J.~D.}\ \bibnamefont {Whittaker}}, \bibinfo {author}
  {\bibfnamefont {J.}~\bibnamefont {Yao}}, \bibinfo {author} {\bibfnamefont
  {A.~W.}\ \bibnamefont {Sandvik}},\ and\ \bibinfo {author} {\bibfnamefont
  {M.~H.}\ \bibnamefont {Amin}},\ }\bibfield  {title} {\bibinfo {title}
  {Quantum critical dynamics in a 5000-qubit programmable spin glass},\ }\href
  {https://arxiv.org/abs/2207.13800} {\bibfield  {journal} {\bibinfo  {journal}
  {arXiv:2207.13800 [quant-ph]}\ } (\bibinfo {year} {2022})}\BibitemShut
  {NoStop}%
\bibitem [{\citenamefont {Carleo}\ \emph {et~al.}(2009)\citenamefont {Carleo},
  \citenamefont {Tarzia},\ and\ \citenamefont
  {Zamponi}}]{PhysRevLett.103.215302}%
  \BibitemOpen
  \bibfield  {author} {\bibinfo {author} {\bibfnamefont {G.}~\bibnamefont
  {Carleo}}, \bibinfo {author} {\bibfnamefont {M.}~\bibnamefont {Tarzia}},\
  and\ \bibinfo {author} {\bibfnamefont {F.}~\bibnamefont {Zamponi}},\
  }\bibfield  {title} {\bibinfo {title} {{Bose-Einstein Condensation in Quantum
  Glasses}},\ }\href {https://doi.org/10.1103/PhysRevLett.103.215302}
  {\bibfield  {journal} {\bibinfo  {journal} {Phys. Rev. Lett.}\ }\textbf
  {\bibinfo {volume} {103}},\ \bibinfo {pages} {215302} (\bibinfo {year}
  {2009})}\BibitemShut {NoStop}%
\bibitem [{\citenamefont {Keren}\ \emph {et~al.}(1996)\citenamefont {Keren},
  \citenamefont {Mendels}, \citenamefont {Campbell},\ and\ \citenamefont
  {Lord}}]{keren1996probing}%
  \BibitemOpen
  \bibfield  {author} {\bibinfo {author} {\bibfnamefont {A.}~\bibnamefont
  {Keren}}, \bibinfo {author} {\bibfnamefont {P.}~\bibnamefont {Mendels}},
  \bibinfo {author} {\bibfnamefont {I.~A.}\ \bibnamefont {Campbell}},\ and\
  \bibinfo {author} {\bibfnamefont {J.}~\bibnamefont {Lord}},\ }\bibfield
  {title} {\bibinfo {title} {{Probing the Spin-Spin Dynamical Autocorrelation
  Function in a Spin Glass above ${T}_{g}$ via Muon Spin Relaxation}},\ }\href
  {https://doi.org/10.1103/PhysRevLett.77.1386} {\bibfield  {journal} {\bibinfo
   {journal} {Phys. Rev. Lett.}\ }\textbf {\bibinfo {volume} {77}},\ \bibinfo
  {pages} {1386} (\bibinfo {year} {1996})}\BibitemShut {NoStop}%
\bibitem [{\citenamefont {Malinovsky}\ \emph {et~al.}(1991)\citenamefont
  {Malinovsky}, \citenamefont {Novikov},\ and\ \citenamefont
  {Sokolov}}]{malinovsky1991log}%
  \BibitemOpen
  \bibfield  {author} {\bibinfo {author} {\bibfnamefont {V.~K.}\ \bibnamefont
  {Malinovsky}}, \bibinfo {author} {\bibfnamefont {V.~N.}\ \bibnamefont
  {Novikov}},\ and\ \bibinfo {author} {\bibfnamefont {A.~P.}\ \bibnamefont
  {Sokolov}},\ }\bibfield  {title} {\bibinfo {title} {Log-normal spectrum of
  low-energy vibrational excitations in glasses},\ }\href
  {https://doi.org/10.1016/0375-9601(91)90363-D} {\bibfield  {journal}
  {\bibinfo  {journal} {Phys. Lett. A}\ }\textbf {\bibinfo {volume} {153}},\
  \bibinfo {pages} {63} (\bibinfo {year} {1991})}\BibitemShut {NoStop}%
\bibitem [{\citenamefont {Rozenberg}\ and\ \citenamefont
  {Grempel}(1998)}]{rozenberg1998dynamics}%
  \BibitemOpen
  \bibfield  {author} {\bibinfo {author} {\bibfnamefont {M.~J.}\ \bibnamefont
  {Rozenberg}}\ and\ \bibinfo {author} {\bibfnamefont {D.~R.}\ \bibnamefont
  {Grempel}},\ }\bibfield  {title} {\bibinfo {title} {{Dynamics of the
  Infinite-Range Ising Spin-Glass Model in a Transverse Field}},\ }\href
  {https://doi.org/10.1103/PhysRevLett.81.2550} {\bibfield  {journal} {\bibinfo
   {journal} {Phys. Rev. Lett.}\ }\textbf {\bibinfo {volume} {81}},\ \bibinfo
  {pages} {2550} (\bibinfo {year} {1998})}\BibitemShut {NoStop}%
\bibitem [{\citenamefont {Ma}\ \emph {et~al.}(2018)\citenamefont {Ma},
  \citenamefont {Wang}, \citenamefont {Dong}, \citenamefont {Zhang},
  \citenamefont {Li}, \citenamefont {Zheng}, \citenamefont {Yu}, \citenamefont
  {Wang}, \citenamefont {Che}, \citenamefont {Ran}, \citenamefont {Bao},
  \citenamefont {Cai}, \citenamefont {\ifmmode~\check{C}\else
  \v{C}\fi{}erm\'ak}, \citenamefont {Schneidewind}, \citenamefont {Yano},
  \citenamefont {Gardner}, \citenamefont {Lu}, \citenamefont {Yu},
  \citenamefont {Liu}, \citenamefont {Li}, \citenamefont {Li},\ and\
  \citenamefont {Wen}}]{ma2018spin-glass}%
  \BibitemOpen
  \bibfield  {author} {\bibinfo {author} {\bibfnamefont {Z.}~\bibnamefont
  {Ma}}, \bibinfo {author} {\bibfnamefont {J.}~\bibnamefont {Wang}}, \bibinfo
  {author} {\bibfnamefont {Z.-Y.}\ \bibnamefont {Dong}}, \bibinfo {author}
  {\bibfnamefont {J.}~\bibnamefont {Zhang}}, \bibinfo {author} {\bibfnamefont
  {S.}~\bibnamefont {Li}}, \bibinfo {author} {\bibfnamefont {S.-H.}\
  \bibnamefont {Zheng}}, \bibinfo {author} {\bibfnamefont {Y.}~\bibnamefont
  {Yu}}, \bibinfo {author} {\bibfnamefont {W.}~\bibnamefont {Wang}}, \bibinfo
  {author} {\bibfnamefont {L.}~\bibnamefont {Che}}, \bibinfo {author}
  {\bibfnamefont {K.}~\bibnamefont {Ran}}, \bibinfo {author} {\bibfnamefont
  {S.}~\bibnamefont {Bao}}, \bibinfo {author} {\bibfnamefont {Z.}~\bibnamefont
  {Cai}}, \bibinfo {author} {\bibfnamefont {P.}~\bibnamefont
  {\ifmmode~\check{C}\else \v{C}\fi{}erm\'ak}}, \bibinfo {author}
  {\bibfnamefont {A.}~\bibnamefont {Schneidewind}}, \bibinfo {author}
  {\bibfnamefont {S.}~\bibnamefont {Yano}}, \bibinfo {author} {\bibfnamefont
  {J.~S.}\ \bibnamefont {Gardner}}, \bibinfo {author} {\bibfnamefont
  {X.}~\bibnamefont {Lu}}, \bibinfo {author} {\bibfnamefont {S.-L.}\
  \bibnamefont {Yu}}, \bibinfo {author} {\bibfnamefont {J.-M.}\ \bibnamefont
  {Liu}}, \bibinfo {author} {\bibfnamefont {S.}~\bibnamefont {Li}}, \bibinfo
  {author} {\bibfnamefont {J.-X.}\ \bibnamefont {Li}},\ and\ \bibinfo {author}
  {\bibfnamefont {J.}~\bibnamefont {Wen}},\ }\bibfield  {title} {\bibinfo
  {title} {{Spin-Glass Ground State in a Triangular-Lattice Compound
  ${\mathrm{YbZnGaO}}_{4}$}},\ }\href
  {https://doi.org/10.1103/PhysRevLett.120.087201} {\bibfield  {journal}
  {\bibinfo  {journal} {Phys. Rev. Lett.}\ }\textbf {\bibinfo {volume} {120}},\
  \bibinfo {pages} {087201} (\bibinfo {year} {2018})}\BibitemShut {NoStop}%
\bibitem [{\citenamefont {Ritort}\ and\ \citenamefont
  {Sollich}(2003)}]{ritort2003glassy}%
  \BibitemOpen
  \bibfield  {author} {\bibinfo {author} {\bibfnamefont {F.}~\bibnamefont
  {Ritort}}\ and\ \bibinfo {author} {\bibfnamefont {P.}~\bibnamefont
  {Sollich}},\ }\bibfield  {title} {\bibinfo {title} {Glassy dynamics of
  kinetically constrained models},\ }\href
  {https://doi.org/10.1080/0001873031000093582} {\bibfield  {journal} {\bibinfo
   {journal} {Advances in physics}\ }\textbf {\bibinfo {volume} {52}},\
  \bibinfo {pages} {219} (\bibinfo {year} {2003})}\BibitemShut {NoStop}%
\bibitem [{\citenamefont {Signoles}\ \emph {et~al.}(2021)\citenamefont
  {Signoles}, \citenamefont {Franz}, \citenamefont {Ferracini~Alves},
  \citenamefont {G\"arttner}, \citenamefont {Whitlock}, \citenamefont
  {Z\"urn},\ and\ \citenamefont {Weidem\"uller}}]{signoles2021glassy}%
  \BibitemOpen
  \bibfield  {author} {\bibinfo {author} {\bibfnamefont {A.}~\bibnamefont
  {Signoles}}, \bibinfo {author} {\bibfnamefont {T.}~\bibnamefont {Franz}},
  \bibinfo {author} {\bibfnamefont {R.}~\bibnamefont {Ferracini~Alves}},
  \bibinfo {author} {\bibfnamefont {M.}~\bibnamefont {G\"arttner}}, \bibinfo
  {author} {\bibfnamefont {S.}~\bibnamefont {Whitlock}}, \bibinfo {author}
  {\bibfnamefont {G.}~\bibnamefont {Z\"urn}},\ and\ \bibinfo {author}
  {\bibfnamefont {M.}~\bibnamefont {Weidem\"uller}},\ }\bibfield  {title}
  {\bibinfo {title} {{Glassy Dynamics in a Disordered Heisenberg Quantum Spin
  System}},\ }\href {https://doi.org/10.1103/PhysRevX.11.011011} {\bibfield
  {journal} {\bibinfo  {journal} {Phys. Rev. X}\ }\textbf {\bibinfo {volume}
  {11}},\ \bibinfo {pages} {011011} (\bibinfo {year} {2021})}\BibitemShut
  {NoStop}%
\bibitem [{\citenamefont {Senaratne}\ \emph {et~al.}(2022)\citenamefont
  {Senaratne}, \citenamefont {Cavazos-Cavazos}, \citenamefont {Wang},
  \citenamefont {He}, \citenamefont {Chang}, \citenamefont {Kafle},
  \citenamefont {Pu}, \citenamefont {Guan},\ and\ \citenamefont
  {Hulet}}]{senaratneSpin2022}%
  \BibitemOpen
  \bibfield  {author} {\bibinfo {author} {\bibfnamefont {R.}~\bibnamefont
  {Senaratne}}, \bibinfo {author} {\bibfnamefont {D.}~\bibnamefont
  {Cavazos-Cavazos}}, \bibinfo {author} {\bibfnamefont {S.}~\bibnamefont
  {Wang}}, \bibinfo {author} {\bibfnamefont {F.}~\bibnamefont {He}}, \bibinfo
  {author} {\bibfnamefont {Y.-T.}\ \bibnamefont {Chang}}, \bibinfo {author}
  {\bibfnamefont {A.}~\bibnamefont {Kafle}}, \bibinfo {author} {\bibfnamefont
  {H.}~\bibnamefont {Pu}}, \bibinfo {author} {\bibfnamefont {X.-W.}\
  \bibnamefont {Guan}},\ and\ \bibinfo {author} {\bibfnamefont {R.~G.}\
  \bibnamefont {Hulet}},\ }\bibfield  {title} {\bibinfo {title} {{Spin-charge
  separation in a one-dimensional Fermi gas with tunable interactions}},\
  }\href {https://doi.org/10.1126/science.abn1719} {\bibfield  {journal}
  {\bibinfo  {journal} {Science}\ }\textbf {\bibinfo {volume} {376}},\ \bibinfo
  {pages} {1305} (\bibinfo {year} {2022})}\BibitemShut {NoStop}%
\bibitem [{\citenamefont {Javanmard}\ \emph {et~al.}(2019)\citenamefont
  {Javanmard}, \citenamefont {Bera},\ and\ \citenamefont {Heyl}}]{heyl2019}%
  \BibitemOpen
  \bibfield  {author} {\bibinfo {author} {\bibfnamefont {Y.}~\bibnamefont
  {Javanmard}}, \bibinfo {author} {\bibfnamefont {S.}~\bibnamefont {Bera}},\
  and\ \bibinfo {author} {\bibfnamefont {M.}~\bibnamefont {Heyl}},\ }\bibfield
  {title} {\bibinfo {title} {Accessing eigenstate spin-glass order from reduced
  density matrices},\ }\href {https://doi.org/10.1103/PhysRevB.99.144201}
  {\bibfield  {journal} {\bibinfo  {journal} {Phys. Rev. B}\ }\textbf {\bibinfo
  {volume} {99}},\ \bibinfo {pages} {144201} (\bibinfo {year}
  {2019})}\BibitemShut {NoStop}%
\bibitem [{\citenamefont {Huang}\ \emph {et~al.}(2020)\citenamefont {Huang},
  \citenamefont {Kueng},\ and\ \citenamefont {Preskill}}]{huang2020predicting}%
  \BibitemOpen
  \bibfield  {author} {\bibinfo {author} {\bibfnamefont {H.-Y.}\ \bibnamefont
  {Huang}}, \bibinfo {author} {\bibfnamefont {R.}~\bibnamefont {Kueng}},\ and\
  \bibinfo {author} {\bibfnamefont {J.}~\bibnamefont {Preskill}},\ }\bibfield
  {title} {\bibinfo {title} {{Predicting many properties of a quantum system
  from very few measurements}},\ }\href
  {https://doi.org/10.1038/s41567-020-0932-7} {\bibfield  {journal} {\bibinfo
  {journal} {Nature Phys.}\ }\textbf {\bibinfo {volume} {16}},\ \bibinfo
  {pages} {1050} (\bibinfo {year} {2020})}\BibitemShut {NoStop}%
\end{thebibliography}%

\newpage


\section*{Supplementary material}

# Supplemental Material for
# "Emergent glassy behavior in a kagome Rydberg atom array"

Zheng Yan,[1] Yan-Cheng Wang,[2] Rhine Samajdar,[3,4] Subir Sachdev,[5,*] and Zi Yang Meng[1,†]

[1]*Department of Physics and HKU-UCAS Joint Institute of Theoretical and Computational Physics, The University of Hong Kong, Pokfulam Road, Hong Kong SAR, China*
[2]*Zhongfa Aviation Institute of Beihang University, Hangzhou 310023, China*
[3]*Department of Physics, Princeton University, Princeton, NJ 08544, USA*
[4]*Princeton Center for Theoretical Science, Princeton University, Princeton, NJ 08544, USA*
[5]*Department of Physics, Harvard University, Cambridge, MA 02138, USA*

## I. QUANTUM MONTE CARLO SCHEMES

Since the glass phase has many local energy minima, its simulation is a difficult problem in both classical and quantum cases. Therefore, here, we employed three different schemes in our quantum Monte Carlo (QMC) simulations to obtain consistent results.

The first one is thermal annealing [1, 2], where the key idea is to set the initial temperature high enough to make the system stay in a paramagnetic (PM) phase. Then, we decrease the simulated temperature slowly to let the system reach the target temperature. Within the adiabatic evolution, the thermal fluctuations can help the system find its ground state.

For the thermal annealing reported in this paper, our QMC simulation is performed with the stochastic series expansion (SSE) QMC algorithm [3–5]. In SSE, the partition function $Z$ is evaluated by a Taylor expansion, and the trace is taken by summing over a complete set of suitably chosen basis states as

$$Z = \sum_{\alpha} \sum_{n=0}^{\infty} \frac{\beta^n}{n!} \langle \alpha | (-H)^n | \alpha \rangle. \quad \text{(S1)}$$

By writing the Hamiltonian as the sum of a set of operators whose matrix elements are easy to calculate, $H = -\sum_i H_i$, and truncating the Taylor expansion at a sufficiently large cutoff $M$, we can further obtain

$$Z = \sum_{\alpha} \sum_{\{i_p\}} \beta^n \frac{(M-n)!}{M!} \left\langle \alpha \left| \prod_{p=1}^{n} H_{i_p} \right| \alpha \right\rangle. \quad \text{(S2)}$$

In order to perform the summation, a Markov chain Monte Carlo procedure can be used to sample the operator sequence $\{i_p\}$ and the trial state $\alpha$. Each step of the update process contains diagonal and off-diagonal updates. In the former, the diagonal operators are inserted into and removed from the operator sequence. Meanwhile, in the latter, the diagonal and off-diagonal operators can be converted into each other. In the annealing process, we slowly increase the cutoff $M$ with a fixed temperature, instead of decreasing the temperature.

The second scheme employed in our study is quenching [6, 7]. The original idea of quenching stems from metallurgy, referring to a process in which materials are cooled quickly. In this way, the memory of the initial state will be retained in the final status of the system. In our simulations, we adopted the spirit of quenching in SSE QMC. The point is to set the initial state in certain configurations (e.g., the nematic state) with a large cutoff $M$, which corresponds to quenching the configuration into a low-temperature state. Therefore, we do not slowly change the temperature but rather freeze the simulation directly at the targeted state. Using this scheme, we obtained the phase boundary between the glass and ordered phases in the phase diagram.

The last scheme that we adopt is the parallel tempering (PT) method, which has been widely used in QMC studies of glassy systems [8, 9]. The main idea behind this method is to exchange the simulated configurations at different parameters according to their sampling weights. In this work, we use the PT method along the parameter lines at fixed $R_b$ and exchange configurations at different $\delta/\Omega$. Note that in the SSE simulation, the weight of sampling a configuration is dependent on $\delta$ in Eq.(S2), which can be rewritten as $Z(\delta) \equiv \sum_c W_c(\delta)$. When several QMC simulations with different parameters $\delta$ are in progression, the two configurations $W_i(\delta)$ and $W_j(\delta')$ from two simulations can be exchanged with the probability

$$P = \min\left(1, \frac{W_i(\delta')W_j(\delta)}{W_i(\delta)W_j(\delta')}\right), \quad \text{(S3)}$$

and PT helps the QMC overcome the local minima in this strategy.

It is worth noting that parallel tempering cannot suitably process the first-order phase transition, so the quenching protocol is still necessary. Utilizing the properties and advantages of the three schemes, we compare the PT (annealing) and quenching data along fixed $R_b$ ($\delta$) lines to map out the entire phase diagram.

---

[*] sachdev@g.harvard.edu
[†] zymeng@hku.hk

## II. EDWARDS-ANDERSON ORDER PARAMETER

As discussed in the main text, we detect glassy behavior in the phase diagram by the Edwards-Anderson order parameter, which is commonly used to quantify the degree of glassy behaviour in disordered systems [10–16]. The original idea in this regard is based on the celebrated replica trick. The density-density correlation function defined as $\overline{\langle n^a(\tau) n^b(\tau') \rangle}$ is nonzero in a glass phase, where $a, b$ are the replica indices labeling the QMC simulations, and $\tau, \tau'$ are different imaginary times. The angular brackets $\langle \cdots \rangle$ denote the average over the QMC simulations in one replica, while $\overline{\langle ... \rangle}$ represents the average of all the $\langle ... \rangle$ among replicas.

Since there is no correlation between two arbitrary replicas, the correlation function above can be rewritten as $\overline{\langle n^a(\tau) n^b(\tau') \rangle} = \overline{\langle n^a \rangle \langle n^b \rangle}$. One can now study a simpler form of the Edwards-Anderson order parameter by considering the diagonal components, i.e., the $a = b$ case [12], where $\overline{\langle n^a \rangle \langle n^b \rangle} = \overline{\langle n \rangle^2}$. Considering the filling and normalization, the Edwards-Anderson order parameter can be rewritten as

$$q_{\text{EA}} = \frac{\sum_{i=1}^{N} \langle n_i - \rho \rangle^2}{N \rho (1 - \rho)}, \quad \text{(S4)}$$

where $N$ is the total number of sites, $\rho$ is the filling, and $n_i = 1 (= 0)$ when the atom on site $i$ is in the Rydberg (ground) state.

An intuitive interpretation of the Edwards-Anderson order parameter is that every atom in the glass is bound to its current state, which is hard to change. Thus, the deviation of each atom from the average filling can characterize the degree of glassiness. In this work, we run 32 independent QMC processes as the replicas. We calculate $q_{\text{EA}}$ first in each QMC process via averaging all the Monte Carlo steps, and then average all the $q_{\text{EA}}$ of different replicas to obtain the final result.

Importantly, the Edwards-Anderson order parameter is sensitive to the QMC method and Monte Carlo steps. In our work, we find that the Edwards-Anderson order parameter decreases to zero quickly via PT simulation. This is because in the PT scheme, different configurations at different parameters can be exchanged with each other, implying that extra dynamics have been added to the system, making the simulation in the glass phase more ergodic. Under these circumstances, the Edwards-Anderson order parameter is not well-defined to characterize the degree of glassiness. Therefore, we use the PT method to first thermalize the QMC configurations and then measure the Edwards-Anderson order parameter via standard QMC. In this work, we use 60000 Monte Carlo steps to obtain each value of $q_{\text{EA}}$.

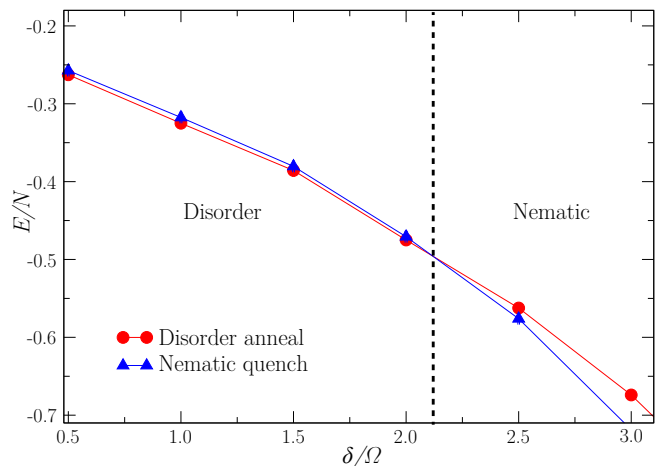

FIG. S1. Weakly first-order phase transition between the disordered and nematic phases. The energy density along the $R_b = 1.9$ path sketched in Fig.1 of the main text, computed using different initial states for the QMC simulations, shows that the nematic and disordered phases are nearly degenerate near the weak first-order phase transition.

## III. QUANTUM PHASE TRANSITIONS TO THE NEMATIC PHASE

In Fig. S1, we illustrate the energy density curves for different initial states and annealing/quenching methods to locate the phase transition between the nematic and glass phases. The simulation scans the cut along the line $R_b = 1.9$ marked in Fig. 1 of the main text. The red curve represents the energy density simulated from random initial configurations with thermal annealing (decreasing the temperature very slowly) [1, 2]. On the other hand, the blue line is simulated from nematic configurations by quenching (initialized at a very low temperature). Deep in the nematic phase, the two energy lines clearly differ. This difference diminishes on approaching the critical point, and the two lines cross and split weakly in the glass phase. Thus, the phase transition between the glass and nematic phases is first-order. We note, however, that the glassy region around $\delta \sim 2$ at $R_b = 1.9$ lies very close to the PM phase with a small Edwards-Anderson order parameter.

Next, we present evidence of the weak first-order phase transition between the nematic and paramagnetic (PM) phases. As shown in Fig. 3(a) in the main text, the energies obtained with both disorder annealing and a nematic quench converge together at about $R_b \simeq 1.3$, with a slight crossing. To further reveal the nature of this phase transition, we calculate the nematic order parameter

$$M_{\text{nematic}} = \left[ \sum_i (n_{i,1} - n_{i,2}) \right]^2 + \left[ \sum_i (n_{i,2} - n_{i,3}) \right]^2 + \left[ \sum_i (n_{i,3} - n_{i,1}) \right]^2, \quad \text{(S5)}$$

where $n_{i,1/2/3}$ represents the density operator on the

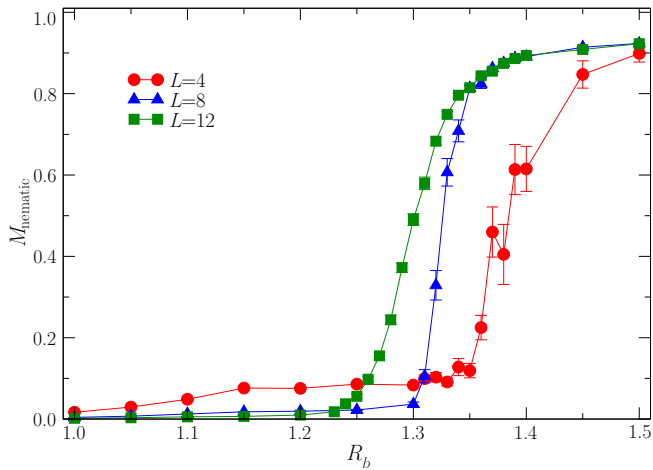

FIG. S2. Order parameter of the nematic phase, $M_{\text{nematic}}$, simulated along the line $\delta/\Omega = 3.3$ for different $R_b$ and $L = \beta$. It is apparent that $M_{\text{nematic}}$ develops a clear discontinuity on increasing the system size $L$.

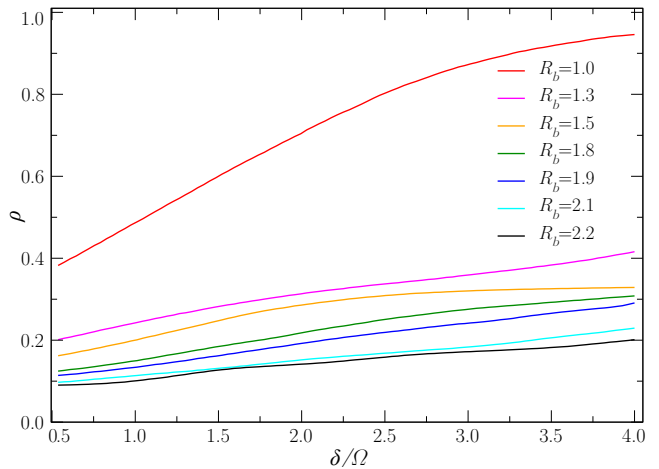

FIG. S3. The filling $\rho$ as obtained via PT. The data presented here is simulated via parallel tempering with $L = \beta = 6$ at different $R_b$.

1/2/3 sublattice of the $i^{\text{th}}$ unit cell. Although the nematic order is homogeneous on one sublattice, it breaks the threefold rotational symmetry among the different sublattices, and can thus be quantified by $M_{\text{nematic}}$.

As shown in Fig. S2, we simulate the nematic order parameter $M_{\text{nematic}}$ with $L = \beta$ along the line $\delta = 3.3$ for different $R_b$. Even though the discontinuity in $M_{\text{nematic}}$ is not immediately apparent for the system sizes studied, previous work by Janke and Villanova [17] has established that there should be a weak first-order phase transition between the nematic and PM phases, which belongs to the $(2+1)$D three-state Potts universality class [18, 19].

## IV. DISTINCTION BETWEEN THE GLASS AND OTHER CANDIDATE PHASES

Previous work on the kagome lattice [18] identified the filling of the disordered phase near the nematic (staggered) phase to be about 1/3 (1/6). Since an even (odd) $\mathbb{Z}_2$ quantum spin liquid (QSL) has a local constraint of two (one) dimers per site of the triangular lattice, which also leads to a filling of 1/3 (1/6) [20], this raises the possibility of the existence of a QSL phase in proximity to these valence bond solids. We observe a similar behavior in our simulation: as shown in Fig. S3, we find that the filling of the disordered region near the nematic (staggered) phase is close to 1/3 (1/6), but it slowly changes as a function of $\delta/\Omega$.

However, the global filling is not a sufficient condition to establish a QSL phase with local symmetry. This requires a measurement that can probe topological properties, such as the loop order parameter, shown in Fig. 3(b) of the main text, which is nearly zero in this region of the phase diagram.

To further demonstrate that the glassy region is indeed different from a $\mathbb{Z}_2$ QSL, we also directly inspect some individual Monte Carlo snapshots to check for local constraints similar to those in the quantum dimer models, i.e., one or two dimers per site. In Fig. S4, these snapshots are drawn in the dimer basis on the triangular lattice, at $\delta/\Omega = 3.3, R_b = 2.05$, for a $L = 8$ system with periodic boundary conditions. Here, a red (blue) bond indicates that the atom positioned at the center of that bond is in the Rydberg (ground) state. All these snapshots display disordered configurations without any obvious local constraint, implying that the system of Rydberg atoms is not well-described by an effective quantum dimer model in this regime, and also conveying the lack of emergent local constraints expected for a $\mathbb{Z}_2$ QSL.

Finally, from prior DMRG studies [18], we know that in the central region between the two VBS phases, there exists a "string phase", which is formed by an order-by-disorder mechanism from an exponentially degenerate manifold of classical configurations with 2/9 Rydberg excitation density. For the sake of completeness, we also use a "string quench" method to compute the energy and compare it to that obtained from PT. In this method, we explicitly set the initial state of the QMC simulations as a

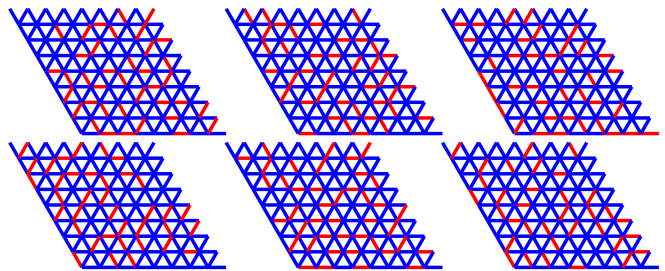

FIG. S4. Snapshots of different QMC processes in the glassy region ($\delta/\Omega = 3.3, R_b = 2.05$), sketched as dimer configurations on the triangular lattice for a $L = 8$ system with periodic boundary conditions. A red (blue) bond represents the presence (absence) of a Rydberg excitation on the atom located at the center of the bond.

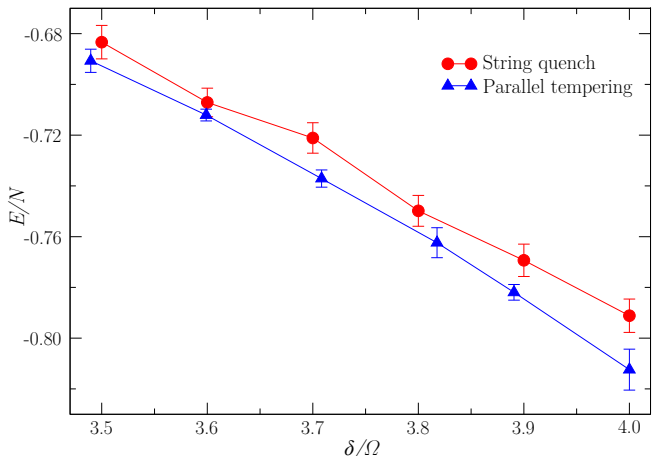

FIG. S5. Energy density along the $R_b = 2.05$ path, computed using different methods for the QMC simulations, showing that the energy of the string state is higher than that of the glass. We set $L = \beta = 6$ in this simulation.

string configuration, using an initial cutoff of $M = 20000$. From Fig. S5, we see that the energy of string phase thus obtained is higher than that of the glass, thereby ruling out the string-ordered solid as a candidate ground state.

[1] J. Brooke, D. Bitko, F. T. Rosenbaum, and G. Aeppli, Quantum annealing of a disordered magnet, Science **284**, 779 (1999).
[2] J. Brooke, T. Rosenbaum, and G. Aeppli, Tunable quantum tunnelling of magnetic domain walls, Nature **413**, 610 (2001).
[3] A. W. Sandvik and J. Kurkijärvi, Quantum Monte Carlo simulation method for spin systems, Phys. Rev. B **43**, 5950 (1991).
[4] A. W. Sandvik, Stochastic series expansion method with operator-loop update, Phys. Rev. B **59**, R14157 (1999).
[5] A. W. Sandvik, Stochastic series expansion method for quantum Ising models with arbitrary interactions, Phys. Rev. E **68**, 056701 (2003).
[6] I. S. Aranson, N. B. Kopnin, and V. M. Vinokur, Dynamics of vortex nucleation by rapid thermal quench, Phys. Rev. B **63**, 184501 (2001).
[7] A. Mitra, Quantum quench dynamics, Annu. Rev. Condens. Matter Phys. **9**, 245 (2018).
[8] K. Hukushima and K. Nemoto, Exchange Monte Carlo method and application to spin glass simulations, J. Phys. Soc. Japan **65**, 1604 (1996).
[9] R. G. Melko, Simulations of quantum XXZ models on two-dimensional frustrated lattices, J. Phys.: Condens. Matter **19**, 145203 (2007).
[10] S. F. Edwards and P. W. Anderson, Theory of spin glasses, J. Phys. F **5**, 965 (1975).
[11] P. M. Richards, Spin-glass order parameter of the random-field Ising model, Phys. Rev. B **30**, 2955 (1984).
[12] K. Binder and A. P. Young, Spin glasses: Experimental facts, theoretical concepts, and open questions, Rev. Mod. Phys. **58**, 801 (1986).
[13] A. Georges, O. Parcollet, and S. Sachdev, Quantum fluctuations of a nearly critical Heisenberg spin glass, Phys. Rev. B **63**, 134406 (2001).
[14] A. Angelone, F. Mezzacapo, and G. Pupillo, Superglass phase of interaction-blockaded gases on a triangular lattice, Phys. Rev. Lett. **116**, 135303 (2016).
[15] Z. Zhou, X.-F. Zhang, F. Pollmann, and Y. You, Fractal quantum phase transitions: Critical phenomena beyond renormalization, arXiv:2105.05851 [cond-mat.str-el] (2021).
[16] A. D. King, J. Raymond, T. Lanting, R. Harris, A. Zucca, F. Altomare, A. J. Berkley, K. Boothby, S. Ejtemaee, C. Enderud, E. Hoskinson, S. Huang, E. Ladizinsky, A. J. R. MacDonald, G. Marsden, R. Molavi, T. Oh, G. Poulin-Lamarre, M. Reis, C. Rich, Y. Sato, N. Tsai, M. Volkmann, J. D. Whittaker, J. Yao, A. W. Sandvik, and M. H. Amin, Quantum critical dynamics in a 5000-qubit programmable spin glass, arXiv:2207.13800 [quant-ph] (2022).
[17] W. Janke and R. Villanova, Three-dimensional 3-state Potts model revisited with new techniques, Nucl. Phys. B **489**, 679 (1997).
[18] R. Samajdar, W. W. Ho, H. Pichler, M. D. Lukin, and S. Sachdev, Quantum phases of Rydberg atoms on a kagome lattice, Proc. Natl. Acad. Sci. U.S.A. **118**, e2015785118 (2021), 2011.12295.
[19] S. Sachdev, *Quantum Phase Transitions* (Cambridge University Press, New York, 2011).
[20] Z. Yan, R. Samajdar, Y.-C. Wang, S. Sachdev, and Z. Y. Meng, Triangular lattice quantum dimer model with variable dimer density, Nat. Commun. **13**, 5799 (2022).



\end{document}